
\ifdefined\ifXeTeX\else
  \newif\ifXeTeX
  \ifdefined\XeTeXversion\XeTeXtrue\else\XeTeXfalse\fi
\fi
\documentclass[pdflatex,twocolumn,sn-nature,referee]{sn-jnl}
\setcitestyle{super,open={},close={}}

\makeatletter
\newcommand{\figlabel}[1]{\def\@currentlabelname{}\label{#1}}
\makeatother

\usepackage{url}
\usepackage{graphicx}
\usepackage{amsmath,amssymb,amsfonts}
\usepackage{amsthm}
\usepackage{booktabs}
\usepackage{tabularx}
\usepackage{array}
\usepackage{ragged2e}
\usepackage{xcolor}
\usepackage{textcomp}
\usepackage{manyfoot}
\usepackage{placeins}

\theoremstyle{thmstyleone}

\theoremstyle{thmstyletwo}

\theoremstyle{thmstylethree}


\pdfoutput=1
\newcommand{\papertitle}{Neural Decoding as Cognitive Inference}
\newsavebox{\mainfigurelegend}
\begin{document}
\title[]{\papertitle}
\author[1,2]{\fnm{Yi} \sur{Guo}}
\author[1,3]{\fnm{Changhong} \sur{Jing}}
\author[1]{\fnm{Yong} \sur{Hu}}
\author[3]{\fnm{Yan} \sur{Liu}}
\author[4]{\fnm{Michael K. P.} \sur{Ng}}
\author[1,2]{\fnm{Shanshan} \sur{Wang}}
\author[1,2]{\fnm{Shuqiang} \sur{Wang}}
\affil[1]{Shenzhen Institutes of Advanced Technology, Chinese Academy of Sciences, Shenzhen, China}
\affil[2]{University of Chinese Academy of Sciences, Beijing, China}
\affil[3]{Department of Computing, The Hong Kong Polytechnic University, Hong Kong}
\affil[4]{Department of Mathematics, Hong Kong Baptist University, Hong Kong}
\abstract{The brain maintains stable cognition despite continuously changing neural activity. How to extract stable cognitive states from variable neural observations remains a central problem in neural decoding. Existing neural decoding methods map neural observations to predefined external labels based on the stimulus-response principle, often capturing recording-specific spurious correlations. Inspired by how the brain infers the world, and specifically by Bayesian brain theory, we recast neural decoding as cognitive inference constrained by brain-intrinsic priors, yielding high-level meta-neural semantic representations. In decoding experiments spanning five neural recording modalities and three cognitive domains (motor, perception and internal mentation), our cognitive inference method reorganized the geometry of neural observation representations, yielding meta-neural semantic representations that exhibited consistent geometric relationships across cognitive tasks and enabled the recovery of stable cognitive states from variable neural observations. Our work provides an account of how the brain maintains relatively stable cognition despite continual changes in the external environment. Cognitive stability is sustained through cognitive inference from changing neural activity, without requiring fixed neural activity patterns.}
\maketitle

\section*{Introduction}

The brain maintains relatively consistent cognitive states despite continuous changes in neural activity~\cite{Walter2019science,gallego2020long}. Neural observations vary with task context, individual differences and recording sessions, whereas cognitive states remain stable. Similarly, audition and vision depend on different sensory inputs and evoke distinct patterns of neural activity, yet the high-level neural representations derived from this activity show systematic semantic consistency \cite{Tyree2023science}. A central problem in neural decoding is therefore how to recover stable cognitive states from changing neural observations~\cite{haynes2006decoding,poldrack2011inferring,kragel2018representation}. Most existing neural decoding methods remain based on the stimulus-response principle. These methods map neural observations onto predefined labels and often capture recording-specific spurious correlations~\cite{mathis2024decoding,peelen2023testing}. If the stability of cognitive states does not depend on fixed patterns of neural activity, rigid stimulus-response mappings are insufficient for decoding stable cognition.

To overcome this limitation, neural decoding needs to reconsider how the brain understands the external world \cite{Dijksterhuis2024science,Can2023science}. According to Bayesian brain theory, the brain does not passively respond to sensory evidence but actively infers the external world from sensory evidence, with brain-intrinsic priors constraining the inference~\cite{knill2004bayesian,krding2004bayesian,friston2010free,berkes2011spontaneous,lange2023bayesian}. The brain's cognitive states are therefore not directly determined by sensory evidence. Even when sensory evidence differs, the brain may infer consistent cognitive states. For neural decoding, this implies that recovering stable cognitive states from changing neural observations requires inference constrained by brain-intrinsic priors.

\begin{figure*}[!tbp]
\centering
\small\linespread{1.0}\selectfont
\refstepcounter{figure}\label{fig:1}
\sbox{\mainfigurelegend}{\parbox{\textwidth}{\noindent\textbf{Fig. \thefigure\quad Overview of the Bayesian brain hierarchical decoding framework.} \textbf{a},~Bayesian brain account linking speech and scene information to latent causes and neural activity measured with EEG, MEG, fMRI, fNIRS and ECoG. \textbf{b},~Hierarchical decoder comprising low-level neural observation $o$, mid-level prior-constrained $u$ and high-level meta-neural semantic $z$ representations, with $z$ supplied to a generative AI model to produce decoded output $y$. \textbf{c},~Task domains: perception (vision and audition), motor (motor imagery) and internal mentation (memory and judgment). \textbf{d},~Cortical semantic-attribution maps showing localized low-level and brain-wide high-level representations. \textbf{e},~Representational geometry of low-level and high-level representations across the three task domains. Open and filled circles distinguish cognitive states; blue and orange distinguish cognitive conditions. Arrows connect states within each condition; greater parallelism across conditions indicates greater cognitive stability. \textbf{f},~Prediction--target correlations for self-generated thought and subjective interpretation, comparing NeuroSTORM with the proposed framework. Boxes show medians and interquartile ranges (IQRs); whiskers extend to the most extreme values within $1.5\times\mathrm{IQR}$ of the box limits, with outliers shown individually. ****$P < 0.0001$. \textbf{g},~Representative open-ended language and visual decoding outputs. Green words mark shared key semantic content in reference and decoded text.\par}}
\includegraphics[width=\textwidth,height=\dimexpr0.96\textheight-\ht\mainfigurelegend-\dp\mainfigurelegend-10pt\relax,keepaspectratio]{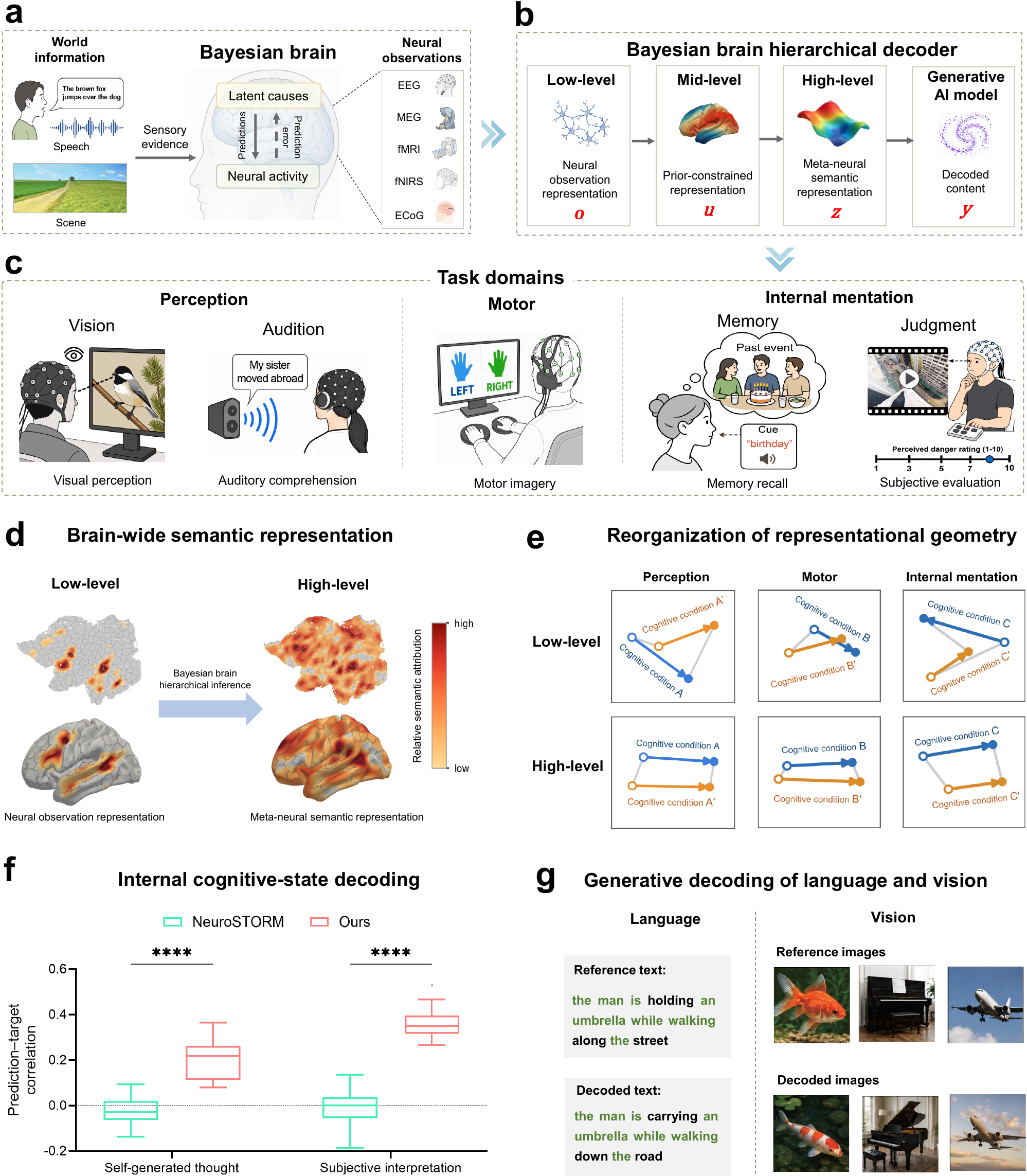}\par
\vspace{6pt}
\usebox{\mainfigurelegend}
\end{figure*}

Building on this view, we recast neural decoding as cognitive inference. Neural observations provide evidence about cognitive states, while brain-intrinsic priors constrain the inference of those states from neural observations. On this basis, we proposed a hierarchical decoding framework inspired by Bayesian brain theory. The framework infers cognitive states from changing neural observations and thereby yields high-level representations that we term meta-neural semantic representations (Fig.~\ref{fig:1}). Within this framework, brain-intrinsic priors may arise from structural constraints of the brain. Previous studies have shown that cortical geometry systematically constrains neural activity throughout the cerebral cortex and that cortical geometric eigenmodes accurately reconstruct the cortex-wide spatial distribution of this activity~\cite{pang2023geometric,Normand2026Geometric,atasoy2016human}. Accordingly, we introduced cortical geometry as a brain-intrinsic prior into cognitive inference. The cortical geometry prior is constructed independently of the cognitive states to be inferred and does not directly encode their specific content. This independence allows us to test whether consistent cognitive states can still be inferred from variable neural observations despite changes in the underlying patterns of neural activity.

We systematically evaluated the framework using five neural modalities: EEG, MEG, ECoG, fNIRS and fMRI. The experiments covered three functional domains, Motor, Perception and Internal mentation, and assessed whether cognitive inference could recover consistent cognitive states from variable neural observations. (1) Compared with neural observation representations, meta-neural semantic representations were more broadly distributed throughout the cortex and exhibited more consistent geometric relations under different cognitive conditions. (2) Meta-neural semantic representations preserved consistent cognitive state relations among different tasks, and decoding performance remained stable despite neural drift over days to months. (3) Neural observations differed markedly between auditory and visual perception, whereas the inferred meta-neural semantic representations remained aligned with the semantic structure of cognitive content. Cognitive state consistency was preserved despite differences in sensory input. (4) Cognitive inference in the proposed framework recovered internally generated cognitive content from neural observations. The framework extended neural decoding from externally evoked content to internal mentation. Overall, cognitive inference in the proposed framework recovered stable cognitive states from variable neural observations. These findings support the view that cognitive stability does not depend on fixed patterns of neural activity but may instead be maintained through continual cognitive inference from variable neural activity.

\section*{Results}\label{sec2}

\subsection*{Brain-intrinsic priors constrain cognitive inference to yield brain-wide meta-neural semantic representations}\label{subsec1}

\begin{figure*}[!tbp]
\centering
\small\linespread{1.0}\selectfont
\refstepcounter{figure}\label{fig:2}
\sbox{\mainfigurelegend}{\parbox{\textwidth}{\noindent\textbf{Fig. \thefigure\quad Brain-intrinsic priors shape brain-wide meta-neural semantic representations.} \textbf{a},~Cortical geometric eigenmodes instantiate brain-intrinsic priors in the hierarchical decoding framework and constrain inference from neural observations to meta-neural semantic representations. \textbf{b},~Vertex-wise correlations between predicted and measured cortical activity for cognitive inference-based encoding and stimulus-response encoding in NSD ($n=8$); evaluation stimuli were held out during model fitting. \textbf{c},~Noise-ceiling-normalized squared prediction correlation for the two encoding models. Boxes show medians and quartiles; whiskers extend to the most extreme values within $1.5\times\mathrm{IQR}$ of the box limits, with outliers shown individually. \textbf{d},~Cortical searchlight RSA maps in NSD show broader cortical correspondence for high-level meta-neural semantic representations than for low-level image features. Color denotes mean Spearman correlation. \textbf{e--h},~Results from SMN4Lang ($n=12$). \textbf{e},~Semantic-attribution maps for three representative participants (Participants A--C). Maps for all 12 participants are shown in Extended Data Fig.~\ref{fig:ed-smn4lang-cortical}. \textbf{f},~Semantic attribution across cortical geometric eigenmodes; the line and shading show the participant mean and s.e.m. The association with eigenmode order was assessed using Spearman's rank correlation. \textbf{g},~Semantic--connectivity coefficients across the whole cortex and seven Yeo networks for CogReader and the proposed framework; values show the participant mean $\pm$ s.e.m. \textbf{h},~Thresholded interregional relations for the two methods, grouped by Yeo network. Edge color denotes cortical geodesic length. A two-sided paired $t$-test was used in \textbf{c}. Asterisks denote statistical significance (*$P < 0.05$, **$P < 0.01$, ***$P < 0.001$ and ****$P < 0.0001$).\par}}
\includegraphics[width=\textwidth,height=\dimexpr0.96\textheight-\ht\mainfigurelegend-\dp\mainfigurelegend-10pt\relax,keepaspectratio]{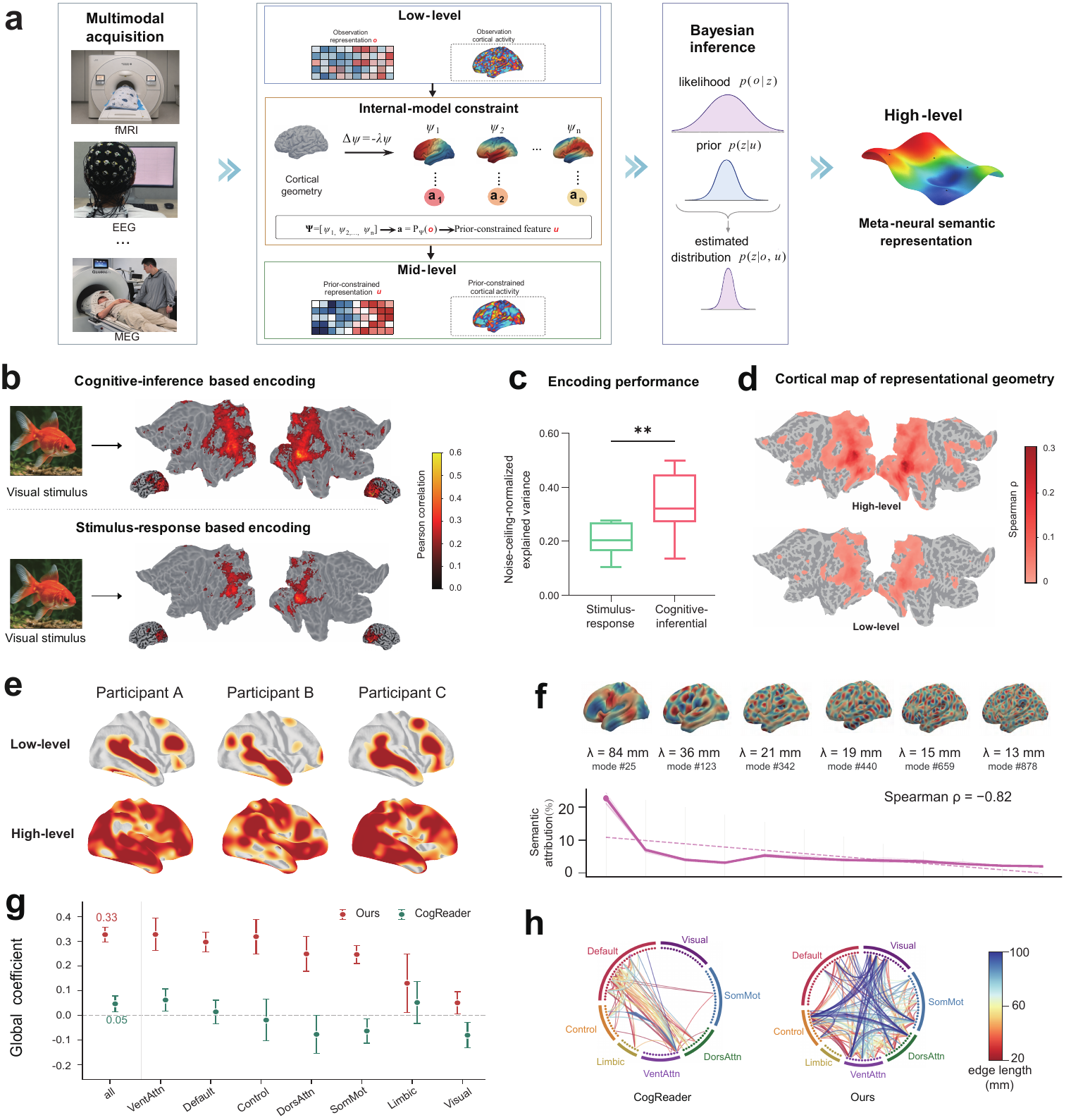}\par
\vspace{6pt}
\usebox{\mainfigurelegend}
\end{figure*}

The Bayesian brain hierarchical decoding framework used cortical geometric eigenmodes to instantiate brain-intrinsic structural priors and inferred meta-neural semantic representations constrained by these priors (Fig.~\ref{fig:2}a).

Cognitive inference-based encoding used the cognitive state representation constrained by cortical geometric eigenmodes to predict cortical activity, whereas stimulus-response-based encoding mapped stimulus semantics to evoked cortical responses. Cognitive inference-based encoding produced significant predictions over a broader cortical range and achieved higher noise-ceiling-normalized squared prediction correlation than stimulus-response-based encoding (Fig.~\ref{fig:2}b--c). These results demonstrate the advantage of the framework in encoding broadly distributed cortical responses constrained by brain-intrinsic priors. This validation was conducted on the Natural Scenes Dataset (NSD; $n=8$ participants).

In NSD, cortical searchlight RSA showed broader cortical correspondence for high-level meta-neural semantic representations than for low-level image features (51.59\% versus 28.28\%; Fig.~\ref{fig:2}d). In SMN4Lang, cortical semantic attribution was more broadly distributed for the high-level meta-neural semantic representation than for the low-level neural observation representation (Fig.~\ref{fig:2}e; $n=12$ participants). From a neural-field perspective, attribution of the meta-neural semantic representation concentrated in long-wavelength cortical geometric eigenmodes, peaked at approximately $84\,\mathrm{mm}$ and decreased with eigenmode order (Spearman $\rho=-0.820$; Fig.~\ref{fig:2}f). This scale dependence related the broader semantic range of the representation to large-scale cortical fields.

The whole-cortex semantic--connectivity coefficient was higher for the proposed framework than for CogReader ($0.33$ versus $0.05$), with the same pattern observed within each of the seven Yeo networks (Fig.~\ref{fig:2}g). The corresponding chord diagrams showed denser relations between network pairs and more long-range connections, as reflected by greater cortical geodesic distances (Fig.~\ref{fig:2}h). These results indicate that the meta-neural semantic representation was more closely aligned with distributed brain-network structure than the CogReader representation.

Cognitive inference constrained by cortical geometry yielded brain-wide meta-neural semantic representations with large-scale spatial relationships. Cognitive state information was thus organized under cortical structural constraints extending beyond local neural activity. These findings support brain-intrinsic structure as a relatively stable constraint on cognitive inference from limited neural observations.

\subsection*{Cognitive inference enables generalizable decoding in diverse tasks}\label{subsec2}

We evaluated the generalizability of discriminative decoding based on meta-neural semantic representations across tasks spanning motor, perception and internal mentation. The evaluation considered representational geometry, within-participant decoding accuracy and zero-shot decoding in unseen participants.

\begin{figure*}[!tbp]
\centering
\small\linespread{1.0}\selectfont
\refstepcounter{figure}\label{fig:3}
\sbox{\mainfigurelegend}{\parbox{\textwidth}{\noindent\textbf{Fig. \thefigure\quad Meta-neural semantic representations exhibit consistent geometric relationships and improved decoding performance across diverse tasks.} \textbf{a},~Three-dimensional MDS projections of the low-level neural observation and high-level meta-neural semantic representations in motor, perception and internal mentation, using the Motor Imagery, NSD and Spacetop datasets, respectively. Open and filled circles distinguish cognitive states; blue and orange distinguish cognitive conditions. Arrows connect cognitive states within each condition, and pale lines connect corresponding states across conditions. Greater parallelism across conditions indicates greater cognitive stability. \textbf{b},~Cross-condition generalization performance (CCGP) and parallelism score (PS) for the low-level and high-level representations. CCGP measures the generalization of cognitive state decoding across conditions, whereas PS measures the parallelism of the corresponding state-difference directions. Violin plots show the distributions of both metrics. \textbf{c},~ Within-participant decoding accuracy for LaBraM, BrainOmni, CBraMod and the proposed framework across BCIC, FACED, SEED-V and Motor Imagery ($n=9$, 123, 16 and 10 participants, respectively). The datasets comprised four motor-imagery classes, three affective states, five affective states and two movement states, respectively. \textbf{d},~ Within-participant confusion matrices for the proposed framework; rows denote true classes, columns predicted classes and cells row-normalized trial percentages. \textbf{e},~ Zero-shot decoding accuracy for participants excluded from model training. In \textbf{c} and \textbf{e}, boxes show medians and quartiles; whiskers extend to the most extreme values within $1.5\times\mathrm{IQR}$ of the box limits, with outliers shown individually. Points denote individual participants. One-way repeated-measures ANOVA with Greenhouse--Geisser correction followed by \v{S}id\'{a}k-adjusted multiple comparisons was used in \textbf{c} and \textbf{e}. Asterisks denote statistical significance (*$P < 0.05$, **$P < 0.01$, ***$P < 0.001$ and ****$P < 0.0001$).\par}}
\includegraphics[width=\textwidth,height=\dimexpr0.96\textheight-\ht\mainfigurelegend-\dp\mainfigurelegend-10pt\relax,keepaspectratio]{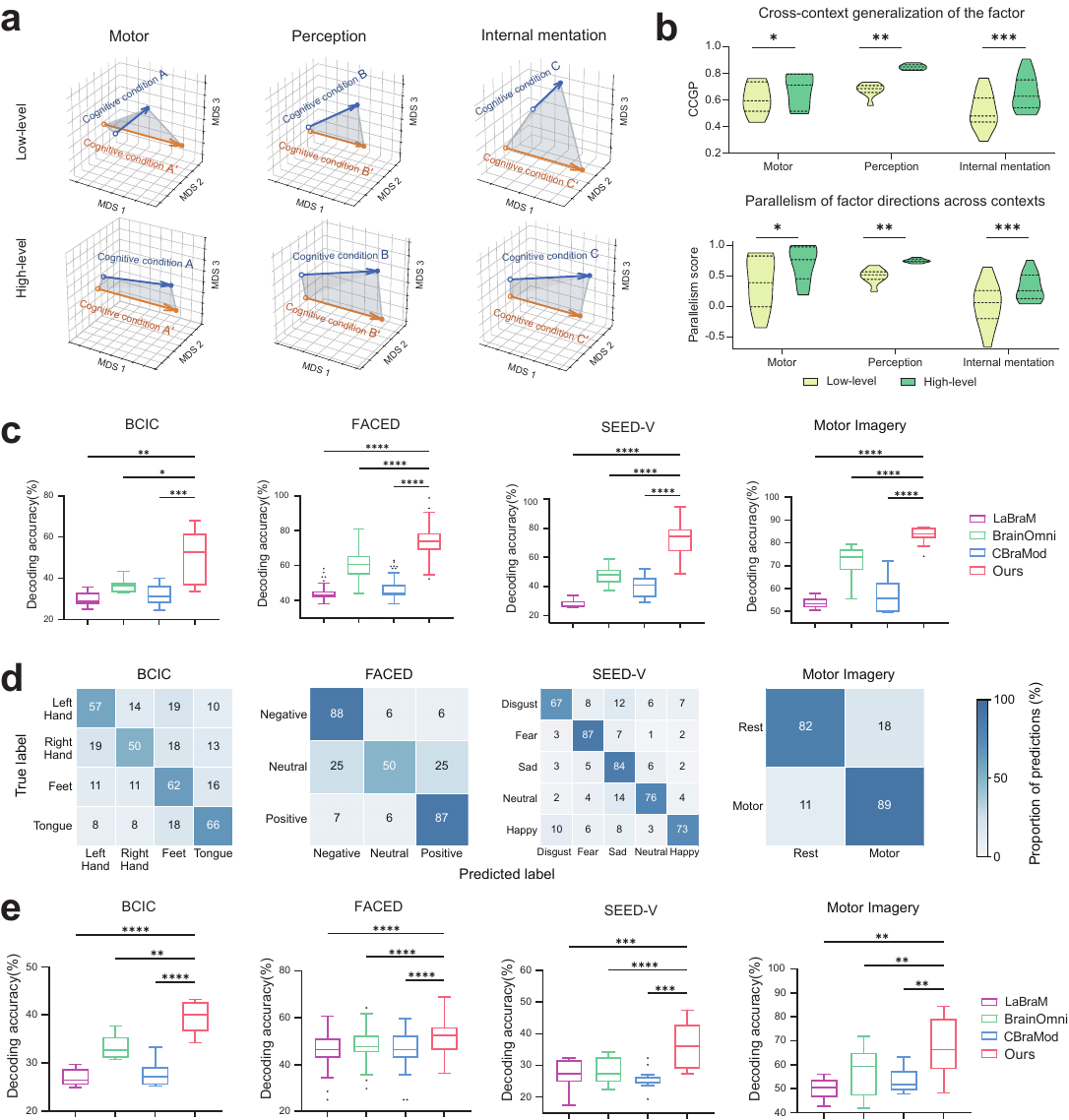}\par
\vspace{6pt}
\usebox{\mainfigurelegend}
\end{figure*}

Three-dimensional MDS projections showed that state-difference directions were more consistent across cognitive conditions at the high level across motor, perception and internal mentation (Fig.~\ref{fig:3}a). Quantitative analyses further showed greater cross-condition generalization performance and stronger parallelism of factor directions at the high level across the three domains (Fig.~\ref{fig:3}b). Complementary analyses showed lower within-state dispersion, greater between-state centroid distance and higher category-structure RSA at the high level (Extended Data Fig.~\ref{fig:ed-clustering}). These results reveal a consistent geometric reorganization of the meta-neural semantic representation, in which task-relevant structure was preserved more consistently across cognitive conditions.

Within-participant decoding in BCIC, FACED, SEED-V and Motor Imagery covered motor-imagery, affective-state and movement-state decoding~\cite{brunner2008bcic2a,chen2023faced,liu2022seedv,lioi2020simultaneous}. Among the four evaluated methods, the proposed framework achieved the highest participant-averaged accuracy in all four datasets compared with LaBraM, BrainOmni and CBraMod (Fig.~\ref{fig:3}c)~\cite{jiang2024labram,xiao2025brainomni,wang2025cbramod}. The box plots showed consistently higher median accuracies for the proposed framework, with participant-level distributions shifted above those of the three comparison methods. The corresponding confusion matrices were dominated by diagonal entries in each dataset (Fig.~\ref{fig:3}d). Thus, the geometric reorganization of the meta-neural semantic representation was accompanied by higher decoding accuracy across multiple tasks.

The proposed framework also achieved the highest mean zero-shot decoding accuracy in each dataset compared with LaBraM, BrainOmni and CBraMod (Fig.~\ref{fig:3}e). All four methods were fitted using only the training participants and applied to unseen participants without participant-specific adaptation. This result extends the performance of the meta-neural semantic representation from within-participant decoding to zero-shot decoding in unseen participants.

Across motor, perception and internal mentation, cognitive inference yielded more consistent relationships among cognitive states despite differences in tasks and neural observations. This geometric consistency was accompanied by improved within-participant and zero-shot decoding. The results suggest that distinct neural activity patterns can support common relationships among cognitive states without requiring identical neural representations across tasks.

\subsection*{Cognitive inference enables stable decoding across time despite neural drift}\label{subsec3}

Meta-neural semantic representations preserved cognitive state relationships and supported cross-session decoding despite neural drift. Repeated recordings spanned 208 days in ECoG Speech and 4 days in AJILE12, both invasive ECoG datasets, and 8 days in non-invasive SHU-MI EEG. These datasets covered speech-command production, natural upper-limb movement and motor imagery.

\begin{figure*}[!tbp]
\centering
\small\linespread{1.0}\selectfont
\refstepcounter{figure}\label{fig:4}
\sbox{\mainfigurelegend}{\parbox{\textwidth}{\noindent\textbf{Fig. \thefigure\quad Meta-neural semantic representations support stable decoding across days to months of neural drift.} \textbf{a, c, e},~ Normalized amplitude profiles of the low-level neural observation and high-level meta-neural semantic representations across sessions in ECoG Speech, AJILE12 and SHU-MI, respectively. \textbf{b, d, f},~ Corresponding cross-session similarity matrices shown on a common scale. \textbf{g},~ Low-level and high-level representational trajectories at D0 and D42 in ECoG Speech, projected onto the first three principal components; temporal profiles and Pearson correlations between D0 and D42 are shown for each component. \textbf{h},~ t-SNE projections at D0 and D42. \textbf{i},~ Silhouette coefficients at D0 and D42 across 50 matched initializations. Boxes show medians and quartiles; whiskers extend to the most extreme values within $1.5\times\mathrm{IQR}$ of the box limits, with outliers shown individually. \textbf{j},~ Representational displacement from D0 across 208 days in ECoG Speech; shading marks follow-ups beyond 200 days. \textbf{k},~ Low-level and high-level representational displacement across the three datasets. \textbf{l},~ Mean ECoG Speech cross-session decoding accuracy for SPaRCNet and the proposed framework at each follow-up; points and error bars show mean $\pm$ s.d. across 20 matched random seeds. \textbf{m},~ Mean cross-session decoding accuracy for SPaRCNet, BrainOmni and the proposed framework across ECoG Speech (six classes), AJILE12 (two classes) and SHU-MI EEG (two classes). Violin plots show distributions across the same 20 matched random seeds; internal boxes indicate medians and interquartile ranges.\par}}
\includegraphics[width=\textwidth,height=\dimexpr\textheight-\ht\mainfigurelegend-\dp\mainfigurelegend-8pt\relax,keepaspectratio]{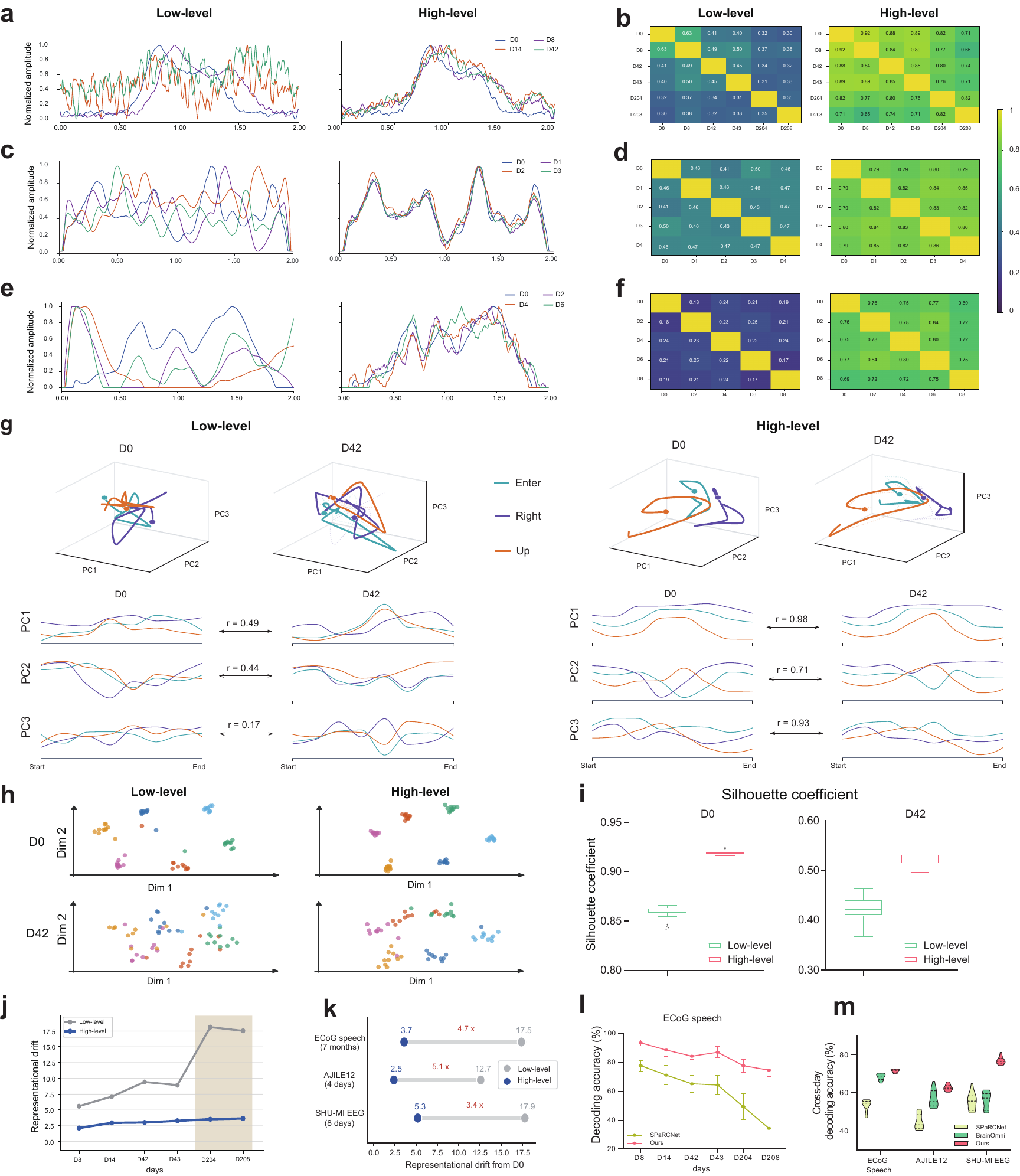}\par
\vspace{6pt}
\usebox{\mainfigurelegend}
\end{figure*}

Normalized amplitude profiles changed markedly between recording sessions in the low-level neural observation representation but remained more consistent in the high-level meta-neural semantic representation (Fig.~\ref{fig:4}a, c, e). The corresponding cross-session similarity matrices showed higher between-session similarity at the high level in ECoG Speech, AJILE12 and SHU-MI (Fig.~\ref{fig:4}b, d, f). Mean similarity increased from $0.393$ to $0.803$ in ECoG Speech, from $0.459$ to $0.819$ in AJILE12, and from $0.214$ to $0.758$ in SHU-MI. The high-level representation therefore preserved condition-specific amplitude patterns as low-level neural observations drifted across sessions.

In ECoG Speech, the high-level representation retained clear state separation in both the reference recording and the 42-day follow-up, whereas the low-level representation showed greater within-class dispersion (Fig.~\ref{fig:4}h). This difference was also reflected in higher silhouette coefficients at the high level (Fig.~\ref{fig:4}i). Moreover, the meta-neural semantic representation preserved more consistent representational trajectory geometry across sessions, whereas the corresponding low-level trajectories changed more substantially between D0 and D42 (Fig.~\ref{fig:4}g). Representational displacement relative to the reference recording remained lower at the high level throughout the 208-day period, including follow-ups more than 200 days after day 0 (shaded interval in Fig.~\ref{fig:4}j). The respective low-level and high-level displacement values were $17.5$ and $3.7$ in ECoG Speech, $12.7$ and $2.5$ in AJILE12, and $17.9$ and $5.3$ in SHU-MI. These values corresponded to reductions by factors of three to five across the three datasets (Fig.~\ref{fig:4}k). This stability was accompanied by higher mean cross-session decoding accuracy. Across 20 matched random seeds, the framework showed higher mean cross-session accuracy than SPaRCNet at every ECoG Speech follow-up and than both SPaRCNet and BrainOmni across all three datasets (Fig.~\ref{fig:4}l--m).

Over days to months of neural drift, meta-neural semantic representations preserved relatively stable relationships among cognitive states and supported cross-session decoding. Cognitive states could therefore be recovered consistently from time-varying neural observations, indicating that stable recovery does not require fixed neural activity patterns.

\subsection*{Cognitive states exhibit semantic consistency across perceptual domains}\label{subsec4}

Meta-neural semantic representations supported open-ended decoding of cognitive content across language and vision.

\begin{figure*}[!tbp]
\centering
\small\linespread{1.0}\selectfont
\refstepcounter{figure}\label{fig:5}
\sbox{\mainfigurelegend}{\parbox{\textwidth}{\noindent\textbf{Fig. \thefigure\quad Cognitive inference enables open-ended decoding of cognitive states across language and visual perceptual domains.} \textbf{a},~ Sentence-level BGEScore between decoded and reference text for the proposed framework, CogReader and a random-output control across Alice, LPPC-fMRI, StudyForrest and LPPC-HK ($n=26$, 112, 20 and 52 participants, respectively). \textbf{b--c},~ Segment-wise BGEScore for Chinese ($n=35$) and French ($n=28$) LPPC-fMRI participants, respectively; lines and error bars show mean $\pm$ s.e.m. \textbf{d},~ Representative decoded passages in English, Chinese, French and German. Green highlighting denotes content semantically consistent with the reference, whereas red highlighting denotes divergent content. \textbf{e--h},~ Within-participant visual decoding in NSD ($n=8$). \textbf{e},~ RSA between image-description semantics and the low-level neural observation or high-level meta-neural semantic representation. \textbf{f--h},~ Comparison with NCC for top-$k$ image retrieval ($k=1$, 5 and 10) (\textbf{f}), viewed-image reconstruction (\textbf{g}) and mental-imagery reconstruction (\textbf{h}). For each retrieval trial, the candidate set comprised the target image and 99 randomly selected non-target images from the 1,000-image test set. \textbf{i},~Representative image-retrieval results. \textbf{j},~Representative image reconstructions. Boxes in \textbf{a} and \textbf{f--h} show medians and quartiles; whiskers extend to the most extreme values within $1.5\times\mathrm{IQR}$ of the box limits, with outliers shown individually. The violin plot in \textbf{e} shows participant-level distributions with medians and quartiles. Two-tailed paired $t$-tests were used in \textbf{a} against CogReader and in \textbf{e--h}; two-tailed Welch's $t$-tests were used in \textbf{a} against the random-output control. Asterisks denote statistical significance (*$P < 0.05$, **$P < 0.01$, ***$P < 0.001$ and ****$P < 0.0001$).\par}}
\includegraphics[width=\textwidth,height=\dimexpr\textheight-\ht\mainfigurelegend-\dp\mainfigurelegend-8pt\relax,keepaspectratio]{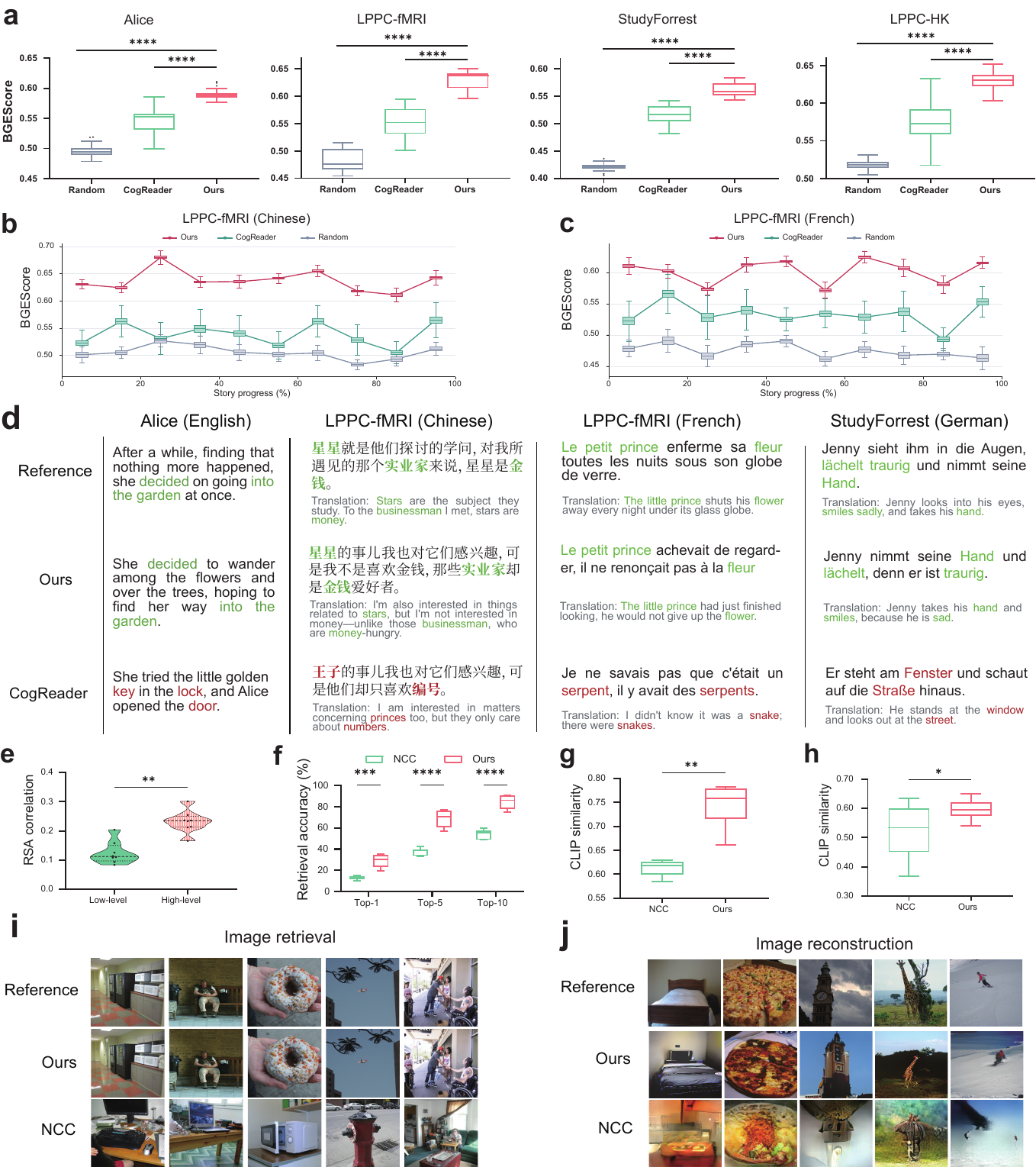}\par
\vspace{6pt}
\usebox{\mainfigurelegend}
\end{figure*}

Language decoding was evaluated in Alice, LPPC-fMRI, StudyForrest and LPPC-HK, covering English, Mandarin Chinese, Cantonese, French and German~\cite{bhattasali2020alice,li2022lppc,hanke2014studyforrest,momenian2024lpphk}. BGEScore measured the sentence-level semantic similarity between decoded and reference text \cite{chen2024bgem3}. The median BGEScore was $0.588$ in Alice, $0.638$ in LPPC-fMRI, $0.559$ in StudyForrest and $0.631$ in LPPC-HK, and the proposed framework outperformed CogReader and the random-output control in all four datasets (all $P<0.0001$; Fig.~\ref{fig:5}a)~\cite{lu2025brain}. This advantage was sustained throughout entire narratives, as illustrated by the Chinese and French stories in LPPC-fMRI (Fig.~\ref{fig:5}b--c). Representative decoded passages retained central entities, actions and event relations from the reference narratives (Fig.~\ref{fig:5}d). These results show that the framework reconstructed text semantics more accurately across multiple languages.

Image decoding from meta-neural semantic representations was evaluated within participants in NSD ($n=8$ participants)~\cite{allen2022nsd}. Representational similarity analysis (RSA) showed a stronger correspondence between the high-level meta-neural semantic representation and image-description semantics than between the low-level neural observation representation and image-description semantics ($P<0.01$; Fig.~\ref{fig:5}e). Compared with Neural Code Conversion (NCC) \cite{wang2025neuralcode}, the proposed framework achieved higher top-1, top-5 and top-10 image-retrieval accuracy ($P<0.001$ for top-1 and $P<0.0001$ for top-5 and top-10; Fig.~\ref{fig:5}f), higher CLIP similarity for reconstruction of viewed images (visual perception) ($P<0.01$; Fig.~\ref{fig:5}g) and mental images (visual imagery) ($P<0.05$; Fig.~\ref{fig:5}h)~\cite{wang2025neuralcode}. Representative retrieval results identified images that matched the target semantics, and representative reconstructions retained the principal objects, scenes and visual content of viewed and imagined targets (Fig.~\ref{fig:5}i--j).

In language and vision, cognitive inference recovered semantic content consistent with the target cognitive states despite differences in sensory inputs and neural activity patterns. This semantic consistency in both perceptual domains supported open-ended decoding, showing that cognitive content could be recovered through cognitive inference across different forms of sensory input.

\subsection*{Cognitive inference extends neural decoding from externally evoked content to internal mentation}\label{subsec5}

Internal cognitive states can arise endogenously during processes such as imagined speech, memory recall and self-generated thought, and can vary between individuals even in response to the same external stimuli. We therefore asked whether these states could be inferred from neural observations using the Bayesian brain hierarchical decoding framework.

\begin{figure*}[!tbp]
\centering
\small\linespread{1.0}\selectfont
\refstepcounter{figure}\label{fig:6}
\sbox{\mainfigurelegend}{\parbox{\textwidth}{\noindent\textbf{Fig. \thefigure\quad Meta-neural semantic representations capture cognitive states in internal mentation.} \textbf{a--c},~Imagined Speech OPM-MEG experiment ($n=6$). \textbf{a},~Reading, imagery and rest phases with corresponding event-related fields. \textbf{b},~Semantic correlation during imagery for BrainOmni and the proposed framework. \textbf{c},~Three-class semantic-category decoding accuracy for imagined content. \textbf{d--g},~Self-generated thought in two public fMRI datasets, comparing NeuroSTORM with the proposed framework. \textbf{d},~State-classification tasks. \textbf{e},~Three-class decoding accuracy in DuPre2016 ($n=31$). \textbf{f},~Two-class decoding accuracy in Lee2021 ($n=26$). \textbf{g},~Correlations between predicted and reported clarity, vividness and valence. \textbf{h--j},~Participant-specific subjective ratings in Spacetop Alignvideo ($n=30$), comparing NeuroSTORM with the proposed framework. \textbf{h},~Shared films and subjective ratings. \textbf{i},~Prediction--rating correlations across five emotional dimensions. \textbf{j},~Residual prediction--rating correlations after removal of the film-specific component shared across participants. \textbf{k},~Representational geometry of happiness for four selected participants, comparing NeuroSTORM with the proposed framework. Arrows represent participants and point towards higher happiness ratings; greater parallelism indicates more consistent representational relationships between happiness states across participants. \textbf{l},~Mean cross-participant parallelism of representational changes associated with happiness ratings ($n=30$). ****$P<0.0001$. \textbf{m},~Pearson correlations between neural representational distance and subjective-rating distance across 136 pairs of 17 participants. \textbf{n},~Participant-level RSA correlations between neural representations and subjective ratings in the same participants. Boxes in \textbf{b}, \textbf{c}, \textbf{e--g} and \textbf{i} show medians and quartiles; whiskers extend to the most extreme values within $1.5\times\mathrm{IQR}$ of the box limits, with outliers shown individually. In \textbf{l}, violin plots show distributions of participant-level mean parallelism, with individual participants shown as black points. Violin plots in \textbf{j} and \textbf{n} show participant-level distributions, with internal lines indicating medians and quartiles. Two-tailed paired $t$-tests were used in \textbf{b}, \textbf{c}, \textbf{e}, \textbf{f} and \textbf{n}. For \textbf{l}, the between-representation difference in mean parallelism was tested using a participant-level leave-one-out jackknife standard error and a two-sided $t$ approximation (29 d.f.), with Holm correction across five emotional dimensions. RSA coefficients in \textbf{n} were Fisher $z$-transformed before testing. Two-way repeated-measures ANOVA followed by \v{S}id\'{a}k-adjusted multiple comparisons was used in \textbf{g}, \textbf{i} and \textbf{j}. Asterisks denote statistical significance (*$P < 0.05$, **$P < 0.01$, ***$P < 0.001$ and ****$P < 0.0001$).\par}}
\includegraphics[width=\textwidth,height=\dimexpr0.96\textheight-\ht\mainfigurelegend-\dp\mainfigurelegend-10pt\relax,keepaspectratio]{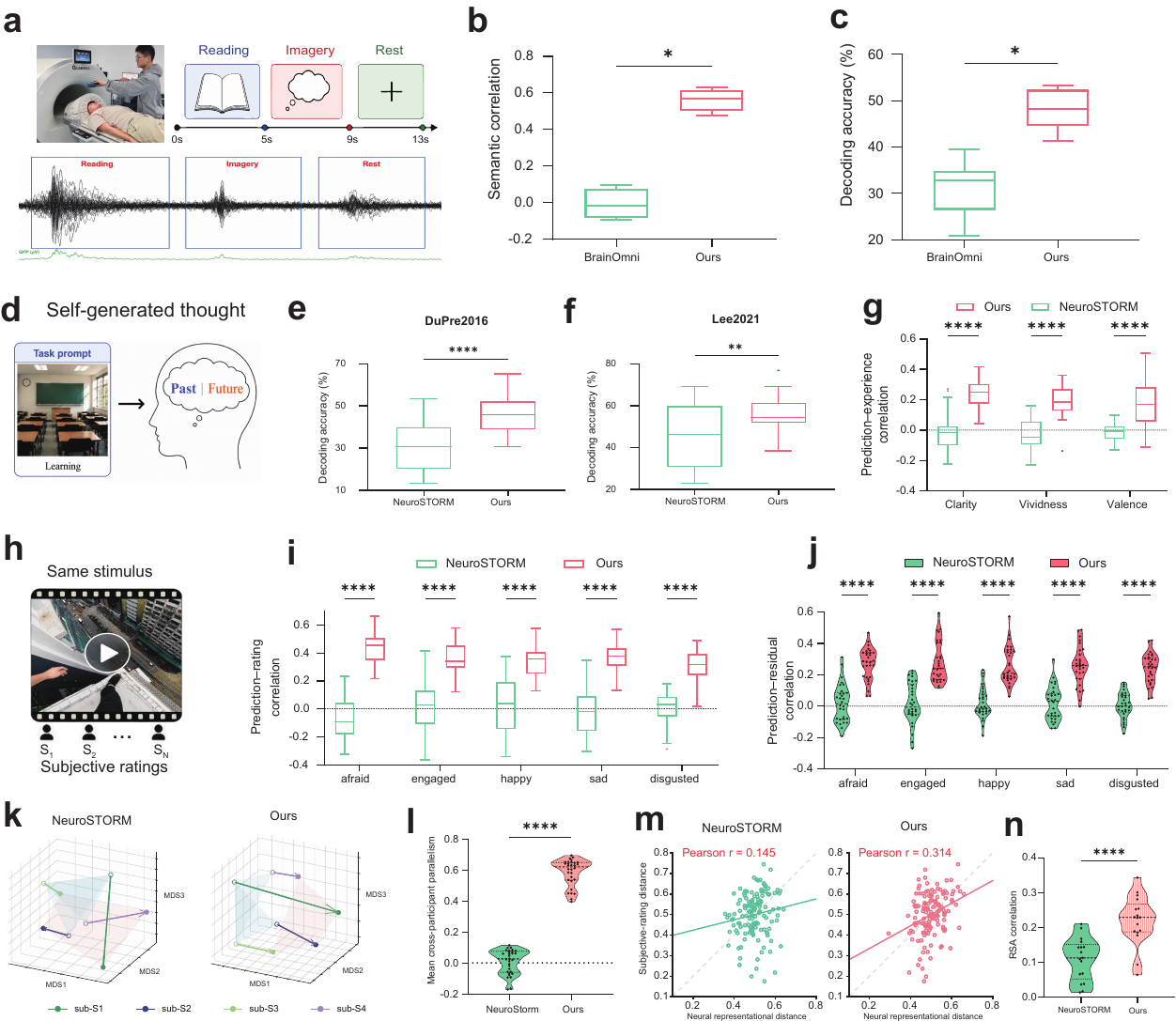}\par
\vspace{6pt}
\usebox{\mainfigurelegend}
\end{figure*}

In the Imagined Speech task, we used an optically pumped magnetometer magnetoencephalography (OPM-MEG) dataset collected for this study ($n=6$ participants). Participants read a text prompt and then imagined its content during a 4-s imagery phase; only neural observations from this phase were used (Fig.~\ref{fig:6}a). The framework achieved a median semantic correlation of $0.567$, which was higher than that obtained with BrainOmni ($P<0.05$; Fig.~\ref{fig:6}b). Median three-class accuracy reached $48\%$, compared with $33\%$ for BrainOmni ($P<0.05$; Fig.~\ref{fig:6}c). These results show that meta-neural semantic representations preserved the semantic content of internally generated cognitive states during imagined speech.

Self-generated thought was assessed in the DuPre2016~\cite{dupre2016autobiographical} and Lee2021 fMRI datasets~\cite{lee2021imagined}. The classification tasks distinguished among self-generated thought states, including autobiographical recollection and imagined future events (Fig.~\ref{fig:6}d). The framework achieved higher state-classification accuracy than NeuroSTORM in DuPre2016 ($P<0.0001$; Fig.~\ref{fig:6}e) and Lee2021 ($P<0.01$; Fig.~\ref{fig:6}f)~\cite{wang2026neurostorm}. Predictions also tracked the clarity of autobiographical recollection and the vividness and valence of imagined future events, with higher prediction--experience correlations than those obtained with NeuroSTORM for all three measures ($P<0.0001$ for all comparisons; Fig.~\ref{fig:6}g). Thus, the framework distinguished self-generated thought states and predicted continuous variation in their reported experience.

Subjective feelings elicited by shared films were assessed in Spacetop Alignvideo. The 30 participants with the most complete film ratings viewed the same videos and rated fear, engagement, happiness, sadness and disgust~\cite{jung2025spacetop} (Fig.~\ref{fig:6}h). Prediction--rating correlations were higher for the proposed framework than for NeuroSTORM in all five dimensions ($P<0.0001$ for all comparisons; Fig.~\ref{fig:6}i). To determine whether this performance captured participant-specific experience rather than only film-related responses shared across participants, the shared film-specific component was removed separately from the observed ratings and model predictions. Correlations between the residual components remained higher than those obtained with NeuroSTORM in all five dimensions ($P<0.0001$ for all comparisons; Fig.~\ref{fig:6}j). The residual correlations therefore captured participant-specific subjective variation after accounting for responses to each film that were shared across participants.

Participant-specific decoding was accompanied by consistent representational relationships between subjective states across individuals. For happiness, these relationships were more parallel across individuals in the proposed framework than in NeuroSTORM (mean parallelism, $0.59$ versus $0.01$; $n=30$; $P<0.0001$; Fig.~\ref{fig:6}k--l). Similar increases were observed for sadness, fear, disgust and engagement ($0.37$--$0.52$ versus $-0.01$--$0.08$; all $P<0.001$). These results indicate that meta-neural semantic representations captured consistent relationships between subjective states across individuals. Across participant pairs, neural representational distance was related to subjective-rating distance for the 17 participants with complete ratings in all five dimensions. The association reached $r=0.314$ for the proposed framework and $r=0.145$ for NeuroSTORM (Fig.~\ref{fig:6}m), indicating that participants with more similar meta-neural semantic representations reported more similar subjective ratings. Within-participant RSA provided a complementary test of the correspondence between representational and rating dissimilarity. Participant-level RSA correlations were higher for the proposed framework than for NeuroSTORM ($P<0.0001$; Fig.~\ref{fig:6}n). Figs.~\ref{fig:ed-rating} and \ref{fig:ed-spacetop-rsa} show corresponding results in additional rating tasks. These results demonstrate that the similarity structure of meta-neural semantic representations aligned with that of subjective evaluations both across and within participants.

Meta-neural semantic representations captured both internally generated cognitive content and distinct participant-specific subjective states elicited by the same stimuli. These findings extend neural decoding to cognitive content generated internally and shaped by individual subjective interpretation, beyond what external stimuli alone specify. Within the framework, such cognitive states were inferred from neural observations under brain-intrinsic priors, rather than identified through a direct mapping between stimuli and neural responses.

\section*{Discussion}

Changes in neural activity patterns do not imply changes in cognitive state~\cite{gallego2020long}. A key question in neural decoding is therefore how to resolve the cognitive states that remain stable in the brain despite continuously changing neural population activity~\cite{haynes2006decoding,poldrack2011inferring}. Existing neural decoding methods, which are largely grounded in the stimulus-response principle, treat predefined external labels as decoding targets and learn statistical correspondences between neural observations and these labels~\cite{mathis2024decoding,peelen2023testing}. When neural activity changes continuously, reliance on such statistical correspondences tends to map changes in neural observations onto changes in cognitive state. These methods thus have difficulty resolving stable cognitive states from changing neural observations. Here, we redefine neural decoding as the process of inferring the cognitive state of the brain from neural observations, and we propose a Bayesian brain hierarchical decoding framework as a computational implementation of this inference. The proposed framework recovers relatively stable cognitive states from changing neural observations, thereby distinguishing changes in neural observations from changes in cognitive states. Accordingly, the framework not only extends the objective of neural decoding but also provides an analytical tool for examining the relationship between changes in neural activity patterns and the stability of cognitive states.

From a neural manifold perspective, cognitive states are embedded within the manifold structure formed by whole-brain neural population activity~\cite{gallego2020long}. Existing neural recording technologies capture only partial observations of this whole-brain activity~\cite{kragel2018representation}. Neural observations are therefore projections of whole-brain neural population activity onto a specific observation space, and they reveal the relationships between cognitive states only incompletely. Our Bayesian brain hierarchical decoding framework treats limited neural observations as evidence and constrains cognitive inference with brain-intrinsic priors. This inference yields meta-neural semantic representations that carry the semantic content of cognitive states. Representational geometry analyses further revealed that cognitive inference was accompanied by a systematic geometric reorganization of the meta-neural semantic representation manifolds (Fig.~\ref{fig:3}a--b; Extended Data Fig.~\ref{fig:ed-clustering}). Cognitive inference altered the relative configuration of the manifolds corresponding to different cognitive states in representational space, such that geometric relationships susceptible to variation in finite neural observations became more stable in the high-level representation. Consequently, the high-level representation preserved the relative relationships among cognitive states even as neural observations changed. At the level of representational geometry, cognitive stability may therefore be understood as the preservation of the relational structure among cognitive states.

Stable cross-session neural decoding does not require neural activity patterns to remain unchanged~\cite{gallego2020long,degenhart2020stabilization,sussillo2016making}. Neural drift substantially alters the neural observations associated with a given cognitive state across sessions, even when the cognitive state itself remains stable. The free-energy principle provides a theoretical perspective on how changes in neural activity can coexist with cognitive stability. Within this framework, free-energy minimization characterizes how the brain maintains relative stability in a changing environment without requiring fixed patterns of neural activity~\cite{friston2010free}. From this perspective, neural drift can be understood as cross-session variation in neural activity that is compatible with stable cognitive states, rather than as evidence that the cognitive states themselves have changed. Our results are consistent with this view. Despite neural drift over days to months, cognitive inference with our method yielded meta-neural semantic representations whose representational geometry remained consistent across sessions (Fig.~\ref{fig:4}). On this basis, the proposed framework maintained stable decoding performance across sessions. Together, these results indicate that stable cross-session neural decoding need not rely on aligning neural observations across sessions, but can instead be achieved by inferring stable cognitive states from changing neural observations.

The cognitive state of the brain is not uniquely determined by external stimuli. Cognitive states also arise from internal processes such as imagination and spontaneous thought, and the same external stimulus gives rise to different subjective interpretations. Decoding based on the stimulus-response principle takes predefined external labels as targets and thereby prescribes stimulus content as the decoding outcome. Such decoding mainly reproduces stimulus information rather than inferring the cognitive state of the brain. In contrast, results from the internal mentation experiments (Fig.~\ref{fig:6}) showed that the meta-neural semantic representations inferred by the proposed framework are not restricted to stimulus content but also reflect internally generated cognitive states and subjective interpretations. Our method therefore extends the scope of neural decoding to internal cognitive processes, including memory, imagination, judgment and subjective interpretation, and provides a computational approach for investigating the cognitive states generated by such processes. This extension broadens the range of information accessible to human-machine interaction and provides a computational foundation for cognitive brain-computer interfaces~\cite{saez2026emerging}.

Human brain structure exhibits pronounced inter-individual variation~\cite{bethlehem2022braincharts}. Because brain-intrinsic priors are shaped by brain structure, this variation implies that cognitive inference in different individuals is constrained by different priors. From this perspective, individual differences in brain structure become a source of individuality in cognitive inference itself. The individualized cortical geometric priors in the proposed framework capture this individuality, and the decoding improvement they bring supports this view (Extended Data Fig.~\ref{fig:ed-individual-geometry}). At the same time, the cognitive states of different individuals are inferred through the same process, and individual differences enter this process through the constraints imposed by each brain's prior. Our method thereby achieves personalized decoding under a common inference principle, constraining cognitive inference with each brain's own structure to recover stable cognitive states from individual neural activity.

In this study, we incorporated constraints imposed by brain structure on neural activity into cognitive inference as brain-intrinsic priors, and instantiated these structural priors using cortical geometric eigenmodes~\cite{pang2023geometric,atasoy2016human}. Future studies could also consider priors defined by neural dynamics and examine how different brain-intrinsic priors shape cognitive stability~\cite{breakspear2017dynamic,demirta2019hierarchical}. For example, temporal priors derived from neural dynamics could allow cognitive inference to directly exploit regularities in the evolution of cognitive states, enabling their formation to be tracked over longer timescales. Overall, our experimental results indicate that, in a continuously changing external environment, the brain need not maintain fixed patterns of neural activity to sustain relatively stable cognitive states, but can instead achieve stability through cognitive inference from changing neural activity.

\section*{Methods}

\subsection*{Bayesian brain hierarchical decoding framework}

Bayesian brain theory motivates hierarchical inference from current neural evidence under constraints imposed by brain-intrinsic priors (Fig.~\ref{fig:1}). We denote the neural observation representation by $o$, the prior-constrained representation by $u$, and the inferred high-level meta-neural semantic representation by $z$. The inference is formulated as
\begin{equation}
\label{eq:bayesian_hierarchy}
p(z\mid o,u)
\propto
p(o\mid z)\,p(z\mid u).
\end{equation}
Here, $p(o\mid z)$ quantifies the compatibility of a candidate
representation $z$ with the current neural observation representation,
whereas $p(z\mid u)$ constrains the candidate representation according to
the brain-intrinsic prior. Their product defines the inference over $z$.

\textbf{Neural observation encoder.}
Modality-specific pretrained encoders transformed neural recordings from EEG, MEG, ECoG, fNIRS and fMRI into the neural observation representation $o$. Grouped spatial-coordinate embeddings and factorized attention accommodated differences in recording length and channel configuration. The resulting $o$ represented the evidence provided by the current neural observation.

\textbf{Brain-intrinsic prior encoder.}
The prior encoder incorporated brain-intrinsic prior information into the current neural representation to produce the prior-constrained representation $u$. In the present implementation, the fixed cortical geometric eigenmode basis instantiated the brain-intrinsic structural prior, and $u$ represented the current neural observation under this structural constraint. Their projection coefficients quantified the expression of the current neural observation within the fixed eigenmode basis, whereas the corresponding eigenvalues indexed spatial scale. These quantities were encoded across spatial scale and time to obtain $u$, which imposed the brain-intrinsic prior constraint during inference of $z$. We denote the eigenmode projection operator by $P_{\Psi}$, such that $\mathbf{a}=P_{\Psi}(o)$ gives the eigenmode coefficient vector associated with neural observation representation $o$.

\textbf{Inference of meta-neural semantic representations.}
To infer the meta-neural semantic representation $z$, we formulated inference under two complementary constraints derived within each recording modality from the same neural observation. The neural observation representation $o$ retained evidence from the current recording, whereas the prior-constrained representation $u$ encoded brain-intrinsic structural constraints. We implemented this Bayesian formulation using a temperature-scaled product of experts~\cite{hinton2002poe}:
\begin{equation}
\label{eq:poe}
p_T(z\mid o,u)
\propto
p(o\mid z)^{1/T_o}
p(z\mid u)^{1/T_u}.
\end{equation}
The positive temperatures $T_o$ and $T_u$ controlled the relative
contributions of the observation and prior-constrained factors,
respectively. At the representation level, the two factors were
implemented as isotropic Gaussian experts with a common scale and means centered at $o$ and $u$, respectively, yielding the
normalized inverse-temperature weights
\begin{equation}
\omega_o
=
\frac{T_o^{-1}}
     {T_o^{-1}+T_u^{-1}},
\qquad
\omega_u
=
\frac{T_u^{-1}}
     {T_o^{-1}+T_u^{-1}},
\end{equation}
and the inferred representation
\begin{equation}
\label{eq:representation_inference}
z
=
\omega_o o+\omega_u u.
\end{equation}
The resulting $z$ constituted the meta-neural semantic representation
used for subsequent decoding.

\textbf{Masked-autoencoding pretraining.}
The neural observation encoder and brain-intrinsic prior encoder were
pretrained independently as masked autoencoders, without access to task
labels. Fifty percent of the temporal and spatial patches were randomly
masked. A reconstruction decoder predicted the masked patches from the
corresponding encoder representation, with reconstruction loss evaluated only
over the masked index set $\Omega$:
\begin{equation}
\mathcal{L}_{\mathrm{MAE}}
=
\frac{1}{|\Omega|}
\sum_{p\in\Omega}
\left\lVert
\hat{x}_p-x_p
\right\rVert_2^2 .
\end{equation}
Here, $x_p$ denotes the target patch at masked position $p$, and
$\hat{x}_p$ denotes its reconstruction. Pretraining was performed
independently for each neural recording modality using only unlabeled neural
recordings from the corresponding training partition.

\subsection*{Cortical geometric eigenmodes}

The current implementation instantiates the brain-intrinsic prior through intrinsic cortical geometry, represented by a fixed eigenmode basis. This geometry constrains the spatial propagation of large-scale neural activity. Its eigenfunctions form an orthogonal basis ordered by spatial scale and fixed across experimental tasks. Modality-specific mappings express this basis in each measurement space. Previous work has shown that these eigenmodes account for the large-scale organization of task activity and resting-state networks~\cite{pang2023geometric,Normand2026Geometric}.

\textbf{Cortical eigenmode construction.} The Laplace--Beltrami operator $\Delta$ was defined on the midthickness triangular cortical surface. Its eigenfunctions satisfy
\begin{equation}
\Delta\phi_k
=
-\lambda_k\phi_k.
\end{equation}
The non-negative eigenvalues were ordered from low to high. Smaller eigenvalues correspond to larger spatial scales. The constant zeroth mode was removed. The remaining $K$ modes were assembled as $\Phi=[\phi_1,\ldots,\phi_K]$.

\textbf{Modality-specific eigenmode mapping.} Cortical eigenmodes were used as a common geometric basis across neural recording modalities. Because different modalities sample neural activity in distinct observation spaces, the cortical eigenmode basis was mapped to each modality according to its spatial sampling geometry, and the resulting modality-specific basis was used to obtain eigenmode coefficient time series from the observed neural signals.

The fMRI eigenproblem was solved separately on both hemispheres. The analysis used the standard 32k\_fs\_LR midthickness surface from the Human Connectome Project~\cite{glasser2013minimal}. Each hemisphere contains 32,492 vertices. The cortical atlas region of interest mask removed the medial wall. A finite-element solver computed the eigenmodes. The analysis retained 1,000 modes from each hemisphere, giving 2,000 modes in total. Cortical time series were projected by least squares to obtain eigenmode coefficient time courses.

For EEG and MEG, dipole sources were constrained to a common cortical source space, and a three-layer boundary-element model was used to compute the forward solution for each modality. For each modality, the resulting lead-field matrix mapped the cortical basis to sensor space.

The forward operator mixes cortical modes into correlated sensor patterns. Singular value decomposition was used to process the mapped basis, and selected components formed a sensor-space basis for estimating coefficient time series. Cortical spatial-scale information was incorporated into the sensor-space representation.

For ECoG, the cortical eigenmodes were evaluated at the contact locations. Dataset-specific contact coordinates were mapped to the standard surface. Contact $j$ was assigned to its nearest cortical vertex $v_j$, giving $\widetilde{\Phi}_{jk}=\phi_k(v_j)$. Tikhonov-regularized least squares estimated the coefficients from the observed contact signals.

For fNIRS, source-detector midpoints projected the surface eigenmodes into channel space. The components were orthogonalized before separate projection of HbO and HbR.

Participant-specific eigenmodes were constructed analogously using each participant's cortical surface. Individual surfaces were registered to the standard 32k\_fs\_LR mesh to maintain vertex correspondence. The resulting participant-specific basis replaced the standard cortical basis in the modality-specific mapping procedures described above, with all subsequent steps unchanged.

\subsection*{Decoding and encoding with meta-neural semantic representations}

\textbf{Decoding from meta-neural semantic representations.} A task-specific linear classification head was applied to the inferred meta-neural semantic representation. For trial $i$, its categorical estimate was
\begin{equation}
q_i(c\mid z_i)
=
\operatorname{softmax}
\!\left(W_{\mathrm{cls}}z_i+b_{\mathrm{cls}}\right)_c.
\end{equation}
For target class $c_i$, the classification head was fitted using cross-entropy:
\begin{equation}
\mathcal{L}_{D}
=
\frac{1}{N}
\sum_{i=1}^{N}
\operatorname{CE}
\!\left(c_i,q_i(\cdot\mid z_i)\right).
\end{equation}
Here, $W_{\mathrm{cls}}$ and $b_{\mathrm{cls}}$ are the classification weights and bias, respectively, $N$ is the number of training trials and $\operatorname{CE}$ denotes cross-entropy. All discriminative accuracy estimates were computed from $q_i(c\mid z_i)$. The pretrained encoders and temperature parameters remained fixed, and only $W_{\mathrm{cls}}$ and $b_{\mathrm{cls}}$ were fitted on the corresponding training partition.

Open-ended language decoding comprised semantic alignment followed by conditional generation (Fig.~\ref{fig:5}). BGE-M3 produced token-level text states that were resampled to a fixed-length semantic sequence~\cite{chen2024bgem3}. A query-based cross-attention transformer (Q-Former) aligned the meta-neural semantic representation with this sequence~\cite{li2023blip2}. Let $\bar{\kappa}$ denote their mean tokenwise cosine similarity. The objective combined reconstruction, cosine-distance and symmetric within-batch contrastive losses:

\begin{equation}
\begin{array}{@{}r@{}l@{}}
\displaystyle
\mathcal{L}_{A}
={}&\displaystyle
\alpha_1\mathcal{L}_{2}
+\alpha_2
\left(1-\bar{\kappa}\right)\\
&\displaystyle {}+\alpha_3\mathcal{L}_{\mathrm{con}}.
\end{array}
\end{equation}
The coefficients $\alpha_1$, $\alpha_2$ and $\alpha_3$ are loss weights. The term $\mathcal{L}_{2}$ denotes reconstruction loss, and $\mathcal{L}_{\mathrm{con}}$ denotes the mean of the two cross-modal retrieval cross-entropies computed between mean-pooled neural and text representations within each batch. Parameters of the alignment model and language model were shared across participants. The same story-level 9:1 split was applied to every participant. Semantic alignment was fitted before the language adapters, with both stages using the same training stories. Test stories were excluded from all model-fitting stages and used only for final generation and evaluation. The aligned sequence was projected into the hidden space of the 3.8-billion-parameter Phi-4-mini-instruct model and provided as a soft prompt~\cite{abouelenin2025phi4mini}. PiSSA initialized the LoRA adapters on the attention projections~\cite{meng2024pissa}. Causal language-model loss was evaluated only at answer tokens. Beam-search decoding included penalties for sequence repetition and repeated $n$-grams. Language-specific text normalization was applied to generated sequences before evaluation.

SD Image Variations v2.0 provided normalized CLIP ViT-L/14 image embeddings~\cite{radford2021clip}. The same cross-attention alignment model mapped the meta-neural semantic representation into this conditioning space and was fitted using the reconstruction, cosine-distance and symmetric within-batch contrastive losses defined above. A frozen latent-diffusion decoder generated images from Gaussian latent noise~\cite{rombach2022ldm}. Sampling used DDIM and classifier-free guidance~\cite{song2021ddim}. Diffusion-model parameters remained fixed. Sampling settings and random seeds were matched within each comparison. NOD trials were used exclusively for visual pretraining. All image-decoding evaluations were conducted within individual NSD participants. Participant-specific models were fitted on each participant's NSD-core training data. Held-out NSD-core trials were used to evaluate viewed-image reconstruction, and the NSD imagery condition was used to evaluate mental-imagery reconstruction.

\textbf{Encoding brain activity through meta-neural semantic representations.} Let $e_i\in\mathbb{R}^{d}$ denote the external semantic representation of stimulus $i$ and $b_i\in\mathbb{R}^{J}$ its cortical GLM beta pattern. A semantic bridge $G$ produced $z_i^{\mathrm{sem}}=G(e_i)$. For training stimuli, the frozen hierarchical model determined the target representation $z_i$ from $b_i$ and its projection constrained by brain-intrinsic priors. Predictions for independent test stimuli were based exclusively on $e_i$.

The target response was decomposed as $b_i=\Phi a_i+\epsilon_i$, where $a_i=\Phi^{+}b_i$ and $\epsilon_i=b_i-\Phi a_i$. Two decoders predicted these components from $z_i^{\mathrm{sem}}$, and the predicted components were combined using positive weights:
\begin{equation}
\hat{b}_i
=
\gamma_{\mathrm{mode}}\Phi D_{\mathrm{mode}}(z_i^{\mathrm{sem}})
+
\gamma_{\mathrm{res}} D_{\mathrm{res}}(z_i^{\mathrm{sem}}).
\end{equation}
Both $\gamma_{\mathrm{mode}}$ and $\gamma_{\mathrm{res}}$ were constrained to be positive. The decoders $D_{\mathrm{mode}}$ and $D_{\mathrm{res}}$ predicted $a_i$ and $\epsilon_i$, respectively. The encoding objective was
\begin{equation}
\begin{array}{@{}r@{}l@{}}
\displaystyle
\mathcal{L}_{E}
={}&\displaystyle
\beta_0\left[1-\cos(z_i^{\mathrm{sem}},z_i)\right]\\
&\displaystyle {}+\beta_1
\left\lVert
D_{\mathrm{mode}}(z_i^{\mathrm{sem}})-a_i
\right\rVert_2^2\\
&\displaystyle {}+
\beta_2
\left\lVert
D_{\mathrm{res}}(z_i^{\mathrm{sem}})-\epsilon_i
\right\rVert_2^2\\
&\displaystyle {}+
\beta_3
\left[
1-\operatorname{corr}
\!\left(\hat{b}_i,b_i\right)
\right].
\end{array}
\end{equation}
Here, $\cos(\cdot,\cdot)$ denotes cosine similarity, $\operatorname{corr}(\cdot,\cdot)$ denotes Pearson correlation and $\beta_0,\ldots,\beta_3$ are loss weights. Caption embeddings were the primary semantic input. Category embeddings and image features served as controls. The bridge, response decoders and positive weights were fitted only on training stimuli.

The stimulus-response encoding baseline directly mapped the external semantic representation $e_i$ to the cortical response $b_i$ using an MLP, without hierarchical inference or cortical-geometric prior constraints. It used the same response targets, training and test stimuli, data partition and optimization protocol as the cognitive-inferential encoding model.

\subsection*{Training and implementation details}

The neural observation and brain-intrinsic prior encoders were independently pretrained as masked autoencoders with a reconstruction objective (masking ratio, 0.5) using AdamW (learning rate, $3\times10^{-4}$) for 60 epochs, with batch sizes of 32 and 64, respectively, and cosine learning-rate decay. The checkpoint with the lowest validation reconstruction loss was retained, after which both encoders were frozen. The positive temperatures $T_o$ and $T_u$ were calibrated separately for each recording modality using the validation partition of the corresponding training data. They controlled the relative strength of the neural observation and prior-constrained factors in the temperature-scaled PoE inference. The temperature pair with the lowest validation loss was retained and fixed for all held-out evaluations. The outputs of the two encoders entered the temperature-scaled product-of-experts (PoE) inference to obtain the meta-neural semantic representation. For language decoding, this representation was aligned with BGE-M3 embeddings through subject-specific layers and a Q-Former using a weighted combination of mean-squared-error, token-wise cosine and contrastive losses for 200 epochs (batch size, 256; AdamW; learning rate, $3\times10^{-4}$; cosine learning-rate decay). For text generation, we then jointly optimized a linear projector and PiSSA-LoRA adapters (rank, 16; dropout, 0.05) on frozen Phi-4-mini using cross-entropy with label smoothing of 0.1 for 15 epochs (batch size, 16; AdamW; learning rates, $3\times10^{-4}$ for the projector and $5\times10^{-5}$ for LoRA; cosine learning-rate decay).

For image decoding, the same representation was aligned with CLIP image embeddings through a visual alignment head using a weighted combination of mean-squared-error, cosine and contrastive losses for 200 epochs (batch size, 64; AdamW; learning rate, $2\times10^{-4}$). The aligned embeddings were decoded into images using a frozen diffusion generator without updating its parameters. All models were trained on two NVIDIA A800 GPUs with 80 GB memory each.

\subsection*{Datasets}

The internally acquired Imagined Speech cohort comprised adults aged 25 to 29 years. Each trial consisted of 5 s of reading, 4 s of speech imagery and 4 s of fixation, with only the imagery epoch included in the main analysis. Neural activity was recorded using a 64-channel single-axis OPM-MEG system (Pyramag Epoch 64, Quanmag Healthcare) in a semi-enclosed magnetically shielded cylinder. A T1-weighted anatomical image was acquired for each participant using a 3.0 T uMR 790 scanner (United Imaging Healthcare). The protocol was approved by the Ethics Committee of the Shenzhen Institute of Advanced Technology (approval no. SIAT-IRB-251015-H1027), and all participants provided written informed consent. All public datasets used in this study, together with their cohort sizes, recording modalities and analysis assignments, are summarized in Extended Data Table~\ref{tab:datasets}. Analysis-specific sample sizes are reported in the corresponding figure legends.

\subsection*{Data preprocessing}

T1-weighted images were processed with FreeSurfer through the ABCD-HCP workflow, which reconstructed native white and pial surfaces and derived the midthickness surface~\cite{feczko2021abcd}. Each surface was registered and resampled to the standard 32k\_fs\_LR mesh. The participant-specific cortical region of interest mask was retained. Participant surfaces supported individual eigenmode construction. The group surface was used for datasets without participant anatomical images.

fMRI data were processed with the ABCD-HCP BIDS pipeline~\cite{feczko2021abcd}, an extension of the Human Connectome Project minimal workflow. Processing comprised denoising, spatial registration, cortical reconstruction and CIFTI generation. Motion and distortion correction followed the information supplied by each dataset. Datasets without field maps entered the workflow without fieldmap correction. Nuisance regression and motion censoring were performed before cortical projection. The final model input was the cortical CIFTI time series on the standard 32k\_fs\_LR mesh. Event windows followed the source annotations. A 6 s hemodynamic offset aligned task events to the blood oxygen level dependent response. Image-encoding analyses used GLM beta maps estimated from the preprocessed time series.

EEG datasets released only as preprocessed epochs were analyzed using the preprocessing provided with the original release. For datasets distributed as continuous recordings, signals were notch filtered at the dataset-specific mains frequency, band-pass filtered between 0.1 and 50 Hz, and resampled to 256 Hz. Independent component analysis was applied where continuous recordings and sufficient channel coverage were available to remove ocular, cardiac and stereotyped myogenic components. Task epochs were defined from the event annotations provided with each dataset. Electrode coordinates were retained for forward modeling.

MEG recordings were notch filtered at the dataset-specific mains frequency, band-pass filtered between 0.1 and 50 Hz, and resampled to 256 Hz. Sensor geometry was retained for forward modeling. For OPM-MEG, sensor positions and orientations were co-registered to the individual anatomical image before source modeling.

fNIRS recordings were processed according to the form in which each dataset was released. Raw intensity data were screened for invalid measurements, converted to optical density, corrected for motion artifacts and band-pass filtered before conversion to hemoglobin concentration changes using the modified Beer--Lambert law. Trial-wise baseline correction was applied where required. Datasets released as processed hemoglobin time series retained the provided preprocessing. Channel correspondence was harmonized across runs before analysis.

ECoG recordings were screened for discontinuities and non-finite samples before filtering. Line noise at the dataset-specific mains frequency and its harmonics was removed, after which signals were band-pass filtered and resampled to 256 Hz. The reference scheme provided with the source release was retained where applicable. Electrode coordinates supplied with each dataset were used for cortical-surface sampling.

\subsection*{Model comparisons}

EEG and MEG comparisons used LaBraM, BrainOmni and CBraMod with their released pretrained weights~\cite{jiang2024labram,xiao2025brainomni,wang2025cbramod}. A task-specific linear head was attached to each backbone. Within-participant comparisons used identical dataset partitions. Zero-shot cross-participant training excluded every sample from the evaluated participant. Cross-participant evaluation froze each pretrained backbone and fitted only its classification head.

Cross-session ECoG and EEG comparisons also included SPaRCNet~\cite{jing2023sparcnet}. Every method was trained on the same reference session and evaluated on the same subsequent-session trials. Internal-state fMRI analyses compared the proposed framework with NeuroSTORM~\cite{wang2026neurostorm}.

Language decoding was compared with CogReader using the same story boundaries~\cite{lu2025brain}. A random control replaced the neural soft prompt with independent standard-normal tokens. The frozen language model and decoding procedure remained unchanged. Image retrieval was compared with NCC on identical NSD candidate sets and test trials within each participant~\cite{wang2025neuralcode}. The fNIRS Transformer comparison used the same within-participant fivefold partitions as the proposed framework~\cite{wang2022fnirst}.

Shared cohorts used identical trial definitions, label vocabularies and test samples across methods. Each released model retained its native backbone and input representation. Data partitions, optimization budgets and evaluation metrics were fixed before test evaluation.

\subsection*{Quantitative analyses}

\textbf{Evaluation protocol.} Training and test data were separated at the analysis-specific grouping level. Within-participant classification used run-blocked folds for datasets with multiple runs and trial-blocked folds otherwise. Zero-shot cross-participant evaluation used disjoint participant sets. Language datasets used a 9:1 story-level split shared across participants, with nine parts for training and one for testing. Held-out stories were excluded from model fitting and used only for final generation and scoring. Visual decoding analyses used separate stimuli for model fitting and evaluation. For cross-session evaluation, models were trained on the reference session and applied to all subsequent sessions without refitting.

\textbf{Cortical semantic attribution.} The semantic-attribution maps in Fig.~\ref{fig:2}e were computed on independent SMN4Lang test data. Attribution for the neural observation representation (low-level) was defined as the square root of non-negative cross-validated vertex-to-semantic-embedding $R^2$ after surface smoothing. Attribution for the meta-neural semantic representation (high-level) was obtained from cross-validated eigenmode-to-semantic-embedding $R^2$, with squared eigenmode amplitudes used to project these values onto the cortical surface. The two attribution maps were independently rank-normalized for visualization. The displayed percentages indicate the proportion of cortical surface vertices represented in each attribution map.

Surface-searchlight RSA compared cortical coverage of low-level and high-level reference geometries in eight NSD participants, using 907--1,000 Shared1000 images per participant and repetition-averaged fMRI responses at 59,412 fsLR32k cortical vertices. Searchlights included vertices within 10 mm along template midthickness mesh edges, within the same hemisphere and excluding the medial wall. Local brain RDMs contained pairwise correlation distances ($1-r$, Pearson correlation) between image-response patterns.

Reference RDMs used 768-dimensional CLIP ViT-L/14 embeddings or 1,024-dimensional meta-neural semantic representations predicted by a participant-specific image-to-representation bridge. Its 2,944 inputs combined CLIP ViT-L/14 embeddings, mean-pooled ViT-bigG/14 tokens and 512 training-fitted ConvNeXt principal components. A 2,944--512--1,024 network with GELU and dropout 0.1 predicted fused posterior means inferred from non-test fMRI. Training used mean-squared error, AdamW (learning rate $3\times10^{-4}$, weight decay 0.01, batch size 512) and a 90:10 fitting--validation split of non-shared images. Validation RDM correlation selected the checkpoint over at most 120 epochs, with checks every five epochs and stopping after 25 epochs without improvement. The configuration was selected on sub-02 validation data and fixed across participants; Shared1000 data were excluded from fitting and validation.

Searchlight RSA correlated the upper triangles of brain and reference RDMs using Spearman correlation. Exact two-sided sign-flip tests across participants (256 assignments) were followed by Benjamini--Hochberg correction jointly over both references and all vertices. Coverage comprised vertices with mean $\rho>0.01$ and $q<0.05$, weighted by template vertex area. Paired high-minus-low RSA differences were tested similarly, with correction across vertices as a separate family.

The contribution of each eigenmode band or latent token was quantified by the decrease in decoding performance after occlusion. For eigenmode band $g$ and latent token $m$, the occlusion effects were
\begin{equation}
\delta_g=s_{\mathrm{full}}-s^{\mathrm{occ}}_g,
\qquad
\delta_m=s_{\mathrm{full}}-s^{\mathrm{occ}}_m,
\end{equation}
where $s_{\mathrm{full}}$ denotes the score obtained with the complete representation, and $s^{\mathrm{occ}}_g$ and $s^{\mathrm{occ}}_m$ denote the corresponding scores after occluding eigenmode band $g$ and latent token $m$, respectively. The score was BGEScore for language decoding and CLIP cosine similarity for visual decoding. Modes were grouped into non-overlapping bands in ascending eigenvalue order. Their representative wavelength was
\begin{equation}
\ell_g=\frac{2\pi}{\sqrt{\bar{\lambda}_g}}.
\end{equation}
The value $\bar{\lambda}_g$ is the mean eigenvalue in band $g$.
The final spatial-attention layer supplied the normalized token-to-mode weight $A_{mk}$. This weight was projected through the cortical basis as $\widetilde{A}_m(v)=\sum_k A_{mk}\phi_k(v)$. The participant-level cortical attribution map was
\begin{equation}
\mathcal{M}(v)
=
\sum_m
\max(\delta_m,0)
\lvert \widetilde{A}_m(v)\rvert.
\end{equation}
The resulting attribution map was non-negative by construction and was normalized before group aggregation.

Cortical attribution maps were averaged within the Schaefer atlas~\cite{schaefer2018local}. Parcels were assigned to the seven Yeo networks~\cite{yeo2011organization}. Inspired by activity-flow mapping~\cite{Cole2016ActivityFlow}, we estimated participant-specific functional-connectivity weights from SMN4Lang task-fMRI using pairwise Pearson correlations between parcel time series. Each attribution map was standardized across parcels, and the diagonal of the resulting connectivity matrix $W$ was set to zero. For target parcel $j$, connectivity-weighted semantic attribution was
\begin{equation}
\hat{s}_j
=
\sum_{\ell\ne j}W_{\ell j}s_\ell.
\end{equation}
Here, $s_\ell$ denotes semantic attribution at source parcel $\ell$. The semantic--connectivity alignment coefficient was
\begin{equation}
\label{eq:semantic_connectivity}
\rho
=
\operatorname{corr}_{j}\!\left(\hat{s}_{j},s_{j}\right).
\end{equation}
Here, $\operatorname{corr}_{j}$ denotes Pearson correlation across target parcels $j$. The same calculation was repeated within each Yeo network. For Fig.~\ref{fig:2}g, alignment coefficients for the proposed framework and CogReader were compared descriptively. For Fig.~\ref{fig:2}h, interregional relations were displayed for both methods. Cross-network relations joined parcels assigned to different Yeo networks. Relations were classified as long-distance when their parcel-centroid geodesic distance exceeded the 75th percentile.

\textbf{Representational geometry analyses.} Analyses crossed two cognitive states with two cognitive conditions: rest versus right-hand imagery with versus without EEG neurofeedback; inanimate versus animate images across small versus large real-world sizes; and low versus high engagement across low versus high sadness, defined by participant-specific rating medians.

Three-dimensional MDS visualizations used centroids of the four state--condition combinations from centered, isotropically scaled representations. Projection onto an orthonormal basis spanning their state- and condition-difference vectors preserved pairwise Euclidean distances; coordinates were normalized by mean inter-centroid distance. Each representation was embedded separately, with quantitative metrics computed in full feature space.

CCGP used $L_2$-regularized logistic regression ($C=0.01$) to classify states in one cognitive condition after training in the other. Feature standardization was fitted on the training condition. Accuracy was averaged over both transfer directions and 20 resamples, balancing all four state--condition combinations without replacement. PS was the cosine similarity between the two condition-specific state-difference vectors, each obtained by subtracting one state centroid from the other in a consistent order. High-minus-low differences in participant-level CCGP and PS were tested using exact two-sided paired sign flips of the mean difference.

\textbf{Representational structure and cross-session stability.} Frozen representations were standardized before t-SNE. Principal-component reduction was performed before the two-dimensional embedding. For Fig.~\ref{fig:4}h--i, t-SNE was performed with 50 random initializations. A silhouette coefficient was calculated for each trial and averaged across trials. For Fig.~\ref{fig:4}i, the same coefficient was calculated after balanced sampling of ten trials per class within each ECoG Speech session. Linear-probe accuracy was used to quantify class separability.

Cosine similarity measured direct alignment between neural and semantic features. Representational similarity analysis used $d_{ij}=1-\operatorname{corr}(h_i,h_j)$, where $h_i$ is the representation of item $i$ and $\operatorname{corr}$ denotes Pearson correlation. The RSA coefficient was the Spearman correlation between the upper triangles of two dissimilarity matrices. Group analyses used Fisher-transformed coefficients. Normalized representational-amplitude profiles, cross-session similarity matrices and population-vector displacement quantified cross-session stability. ECoG Speech models were trained on the reference session and applied to every follow-up session. AJILE12 ECoG and SHU-MI used the same reference-session protocol.

\textbf{Internal cognitive state analyses.} Imagined Speech analyses used only the 4 s OPM-MEG imagery epoch. Semantic correlation was quantified using category-level RSA. Trial-level neural representations were averaged within each category to obtain neural centroids. BGE embeddings of the corresponding sentences were averaged within the same categories to obtain semantic centroids. Pairwise distances were calculated separately among neural and semantic centroids, and the Spearman correlation between the corresponding upper-triangular distance values yielded the RSA coefficient. A classifier fitted to imagery epochs yielded content accuracy. DuPre2016 and Lee2021 used their released task labels for self-generated condition decoding. Their trial-level clarity and vividness scores supplied continuous targets. Valence supplied a further continuous target.

Spacetop Alignvideo analyses in Fig.~\ref{fig:6}h--j included participants with complete ratings across all five dimensions for the common set of videos. A separate regression model was fitted for each dimension. Direct performance was the within-participant correlation between predicted and observed ratings. Video-specific consensus was estimated from training participants and subtracted from both the test-participant rating and the model prediction. The correlation between these residuals quantified participant-specific interpretation. Participant-level correlations were Fisher transformed for inference.

The RSA analyses in Fig.~\ref{fig:6}m--n used participants with complete ratings across all five dimensions for the same set of videos. The dimensions were happy, sad, afraid, disgusted and engaged. The item index $i$ identifies a video in this analysis. Rating vectors were standardized using parameters estimated from the training participants and are denoted $\widetilde{y}_{si}$. The analyses compared time-pooled representations from NeuroSTORM and the proposed framework.

For the between-participant analysis, each method yielded a representation matrix for each video, with one row per participant. Neural distance between participants $s_1$ and $s_2$ for video $i$ was
\begin{equation}
d_i^{(z)}(s_1,s_2)
=
1-\operatorname{corr}
\!\left(z_{s_1i},z_{s_2i}\right).
\end{equation}
Rating distance for the same participant pair was
\begin{equation}
d_i^{(y)}(s_1,s_2)
=
\left\lVert
\widetilde{y}_{s_1i}
-\widetilde{y}_{s_2i}
\right\rVert_2.
\end{equation}
Each video yielded participant-pair distances of each type. Fractional ranks were calculated over participant pairs within each video and averaged over videos:
\begin{equation}
\begin{array}{@{}r@{}l@{}}
\displaystyle
\overline{\mathcal{R}}^{\,(\xi)}(s_1,s_2)
={}&\displaystyle
\frac{1}{N}
\sum_{i=1}^{N}
\mathcal{R}_{i}
\!\left(d_i^{(\xi)}(s_1,s_2)\right),\\
&\displaystyle
\xi\in\{z,y\}.
\end{array}
\end{equation}
Here, $N$ denotes the number of videos in this analysis. The operator $\mathcal{R}_{i}$ maps the participant-pair distances for video $i$ to fractional ranks between zero and one. The reported coefficient was the Pearson correlation between the upper-triangular entries of the two participant-pair fractional-rank matrices. This analysis quantified the association between neural representational distance and rating distance.

For the within-participant analysis, each participant yielded neural and rating dissimilarity matrices across videos. Their upper-triangular entries yielded one Pearson coefficient $r_s$. Coefficients were Fisher transformed as $\zeta_s=\operatorname{arctanh}(r_s)$. The transformed values were averaged across participants. The hyperbolic tangent returned the group coefficient to the correlation scale. The participant was the analysis unit. Group differences used a two-tailed paired $t$-test on Fisher-transformed values.

For both RSA analyses, scaling parameters and model selection were estimated from the training data and held fixed during evaluation. Reported coefficients were then calculated on the corresponding held-out observations.

For Extended Data Fig.~\ref{fig:ed-spacetop-rsa}, Spacetop participants who had ratings for both tasks were included. The tasks were Narratives and Shortvideo. RSA compared meta-neural semantic dissimilarity with rating dissimilarity across the shared stimuli. The participant was the unit of analysis.

\textbf{Structural controls and robustness.} Equal-dimensional Fourier and spherical-harmonic bases served as structural controls. Graph-Laplacian and cortical-eigenmode bases completed the comparison. Every basis used the same encoder and classifier. Data partitions were also fixed. Mode-number analyses changed only the number of retained basis functions. Group-template and participant-specific bases were evaluated on identical trials.

Robustness was assessed by adding zero-mean Gaussian noise scaled by the training-fold channel standard deviation, masking nested proportions of channels and restricting the training set to nested fractions of trials. At each perturbation level, BrainOmni and the proposed model received the same perturbation and sample subset. A separate cross-participant analysis progressively added training data from the target participant while holding the test set fixed. These analyses are reported in Extended Data Fig.~\ref{fig:ed-robustness}.

\textbf{Performance metrics and statistical analysis.} Discriminative decoding was summarized by accuracy. Language generation was evaluated using sentence-level BGE-M3 embedding cosine similarity \cite{chen2024bgem3}, termed BGEScore. Image reconstruction used CLIP image-embedding cosine similarity. Retrieval reported top-1, top-5 and top-10 accuracy in a 100-way task. For each test image, the candidate set comprised the target image and 99 randomly selected non-target images from the 1,000-image test set.

Semantic encoding of brain activity was evaluated on independent test stimuli. Vertex-wise prediction correlation was $r_v=\operatorname{corr}(\{\hat{b}_{i,v}\}_i,\{b_{i,v}\}_i)$. Let $n_v$ denote the repeated-response noise ceiling at vertex $v$. The noise-ceiling-normalized squared prediction correlation was
\begin{equation}
r^2_{\mathrm{norm},v}
=
\frac{r_v^2}{n_v}.
\end{equation}
This measure uses squared test-set Pearson correlation, rather than residual-sum-of-squares $R^2$. Vertices with $n_v\leq0$ were excluded. Encoding coverage was the proportion of vertices with positive prediction correlation and a Benjamini--Hochberg adjusted $P<0.05$. Vertex-wise null distributions used $10^4$ permutations of stimulus identity. Semantic-attribution maps and network edges used the same number of permutations. Spatial correspondence, network concentration and coverage differences used $10^4$ spherical rotations. Rotations were performed separately in each hemisphere and preserved the medial wall. Empirical probabilities used the standard add-one correction.

All hypothesis tests were two-sided. Comparisons among matched methods or structural bases used one-way repeated-measures ANOVA. Each participant provided one matched row of measurements. Analyses involving method and a second repeated factor used two-way repeated-measures ANOVA. Sphericity was not assumed, and the Greenhouse--Geisser correction was applied where appropriate. Pairwise contrasts used \v{S}id\'{a}k-adjusted multiple comparisons. Direct comparisons between two matched conditions used two-tailed paired $t$-tests. Comparisons between language-decoding outputs and the independent Random control used two-tailed Welch's $t$-tests.

Line plots show the mean and s.e.m. when specified. The participant was the biological replicate. Session-level analyses treated the session as the experimental unit only when explicitly stated. Repeated t-SNE initializations and matched random-seed model fits quantified optimization stability. Inferential tests used the participant, session, initialization or random-seed sampling units specified in the corresponding figure legends.

\bibliography{refs}

@article{Normand2026Geometric,
  author  = {Normand, Francis and Gajwani, Mehul and Cao, Trang and Cruddas, Jace and Sangchooli, Arshiya and Oldham, Stuart and Holmes, Alexander and Robinson, Peter A. and Pang, James C. and Fornito, Alex},
  title   = {Geometric constraints on the architecture of mammalian cortical connectomes},
  journal = {Cell},
  volume  = {189},
  pages   = {5283--5303.e17},
  year    = {2026},
  doi     = {10.1016/j.cell.2026.05.048}
}

@article{saez2026emerging,
  author = {Saez, Ignacio},
  title = {The emerging field of cognitive brain--computer interfaces},
  journal = {Trends Cogn. Sci.},
  year = {2026},
  note = {\url{https://doi.org/10.1016/j.tics.2026.06.012}}

}

@article{degenhart2020stabilization,
  author = {Degenhart, Alan D. and Bishop, William E. and Oby, Emily R. and Tyler-Kabara, Elizabeth C. and Chase, Steven M. and Batista, Aaron P. and Yu, Byron M.},
  title = {Stabilization of a brain–computer interface via the alignment of low-dimensional spaces of neural activity},
  year = {2020},
  journal = {Nat. Biomed. Eng.},
  volume = {4},
  number = {7},
  pages = {672--685},
  doi = {10.1038/s41551-020-0542-9}
}

@article{gallego2020long,
  author = {Gallego, Juan A. and Perich, Matthew G. and Chowdhury, Raeed H. and Solla, Sara A. and Miller, Lee E.},
  title = {Long-term stability of cortical population dynamics underlying consistent behavior},
  year = {2020},
  journal = {Nat. Neurosci.},
  volume = {23},
  number = {2},
  pages = {260--270},
  doi = {10.1038/s41593-019-0555-4}
}

@article{sussillo2016making,
  author = {Sussillo, David and Stavisky, Sergey D. and Kao, Jonathan C. and Ryu, Stephen I. and Shenoy, Krishna V.},
  title = {Making brain–machine interfaces robust to future neural variability},
  year = {2016},
  journal = {Nat. Commun.},
  volume = {7},
  number = {1},
  pages = {13749},
  doi = {10.1038/ncomms13749}
}

@article{haynes2006decoding,
  author = {Haynes, John-Dylan and Rees, Geraint},
  title = {Decoding mental states from brain activity in humans},
  journal = {Nat. Rev. Neurosci.},
  year = {2006},
  volume = {7},
  number = {7},
  pages = {523--534},
  doi = {10.1038/nrn1931}
}

@article{poldrack2011inferring,
  author = {Poldrack, Russell A.},
  title = {Inferring mental states from neuroimaging data: From reverse inference to large-scale decoding},
  journal = {Neuron},
  year = {2011},
  volume = {72},
  number = {5},
  pages = {692--697},
  doi = {10.1016/j.neuron.2011.11.001}
}

@article{Cole2016ActivityFlow,
  author  = {Cole, Michael W. and Ito, Takuya and Bassett, Danielle S. and Schultz, Douglas H.},
  title   = {Activity flow over resting-state networks shapes cognitive task activations},
  journal = {Nat. Neurosci.},
  volume  = {19},
  pages   = {1718--1726},
  year    = {2016},
  doi     = {10.1038/nn.4406}
}

@article{mathis2024decoding,
  author = {Mathis, Mackenzie Weygandt and Perez Rotondo, Adriana and Chang, Edward F. and Tolias, Andreas S. and Mathis, Alexander},
  title = {Decoding the brain: From neural representations to mechanistic models},
  journal = {Cell},
  year = {2024},
  volume = {187},
  number = {21},
  pages = {5814--5832},
  doi = {10.1016/j.cell.2024.08.051}
}

@article{peelen2023testing,
  author = {Peelen, Marius V. and Downing, Paul E.},
  title = {Testing cognitive theories with multivariate pattern analysis of neuroimaging data},
  journal = {Nat. Hum. Behav.},
  year = {2023},
  volume = {7},
  number = {9},
  pages = {1430--1441},
  doi = {10.1038/s41562-023-01680-z}
}

@article{kragel2018representation,
  author = {Kragel, Philip A. and Koban, Leonie and Barrett, Lisa Feldman and Wager, Tor D.},
  title = {Representation, pattern information, and brain signatures: from neurons to neuroimaging},
  journal = {Neuron},
  year = {2018},
  volume = {99},
  number = {2},
  pages = {257--273},
  doi = {10.1016/j.neuron.2018.06.009}
}

@article{friston2010free,
  author = {Friston, Karl},
  title = {The free-energy principle: a unified brain theory?},
  year = {2010},
  journal = {Nat. Rev. Neurosci.},
  volume = {11},
  number = {2},
  pages = {127--138},
  doi = {10.1038/nrn2787}
}

@article{knill2004bayesian,
  author = {Knill, David C. and Pouget, Alexandre},
  title = {The {Bayesian} brain: the role of uncertainty in neural coding and computation},
  year = {2004},
  journal = {Trends Neurosci.},
  volume = {27},
  number = {12},
  pages = {712--719},
  doi = {10.1016/j.tins.2004.10.007}
}

@article{berkes2011spontaneous,
  author = {Berkes, Pietro and Orbán, Gergő and Lengyel, Máté and Fiser, József},
  title = {Spontaneous cortical activity reveals hallmarks of an optimal internal model of the environment},
  year = {2011},
  journal = {Science},
  volume = {331},
  number = {6013},
  pages = {83--87},
  doi = {10.1126/science.1195870}
}

@article{krding2004bayesian,
  author = {Körding, Konrad P. and Wolpert, Daniel M.},
  title = {Bayesian integration in sensorimotor learning},
  year = {2004},
  journal = {Nature},
  volume = {427},
  number = {6971},
  pages = {244--247},
  doi = {10.1038/nature02169}
}

@article{lange2023bayesian,
  author = {Lange, Richard D. and Shivkumar, Sabyasachi and Chattoraj, Ankani and Haefner, Ralf M.},
  title = {Bayesian encoding and decoding as distinct perspectives on neural coding},
  year = {2023},
  journal = {Nat. Neurosci.},
  volume = {26},
  number = {12},
  pages = {2063--2072},
  doi = {10.1038/s41593-023-01458-6}
}

@article{demirta2019hierarchical,
  author = {Demirtaş, Murat and Burt, Joshua B. and Helmer, Markus and Ji, Jie Lisa and Adkinson, Brendan D. and Glasser, Matthew F. and Van Essen, David C. and Sotiropoulos, Stamatios N. and Anticevic, Alan and Murray, John D.},
  title = {Hierarchical Heterogeneity across Human Cortex Shapes Large-Scale Neural Dynamics},
  year = {2019},
  journal = {Neuron},
  volume = {101},
  number = {6},
  pages = {1181--1194.e13},
  doi = {10.1016/j.neuron.2019.01.017}
}

@article{pang2023geometric,
  author = {Pang, James C. and Aquino, Kevin M. and Oldehinkel, Marianne and Robinson, Peter A. and Fulcher, Ben D. and Breakspear, Michael and Fornito, Alex},
  title = {Geometric constraints on human brain function},
  year = {2023},
  journal = {Nature},
  volume = {618},
  number = {7965},
  pages = {566--574},
  doi = {10.1038/s41586-023-06098-1}
}

@article{atasoy2016human,
  author = {Atasoy, Selen and Donnelly, Isaac and Pearson, Joel},
  title = {Human brain networks function in connectome-specific harmonic waves},
  year = {2016},
  journal = {Nat. Commun.},
  volume = {7},
  number = {1},
  pages = {10340},
  doi = {10.1038/ncomms10340}
}

@article{breakspear2017dynamic,
  author = {Breakspear, Michael},
  title = {Dynamic models of large-scale brain activity},
  year = {2017},
  journal = {Nat. Neurosci.},
  volume = {20},
  number = {3},
  pages = {340--352},
  doi = {10.1038/nn.4497}
}

@article{simistiraliwicki2023bimodal,
  author = {Simistira Liwicki, Foteini and Gupta, Vibha and Saini, Rajkumar and De, Kanjar and Abid, Nosheen and Rakesh, Sumit and Wellington, Scott and Wilson, Holly and Liwicki, Marcus and Eriksson, Johan},
  title = {Bimodal electroencephalography--functional magnetic resonance imaging dataset for inner-speech recognition},
  journal = {Sci. Data},
  year = {2023},
  volume = {10},
  number = {1},
  pages = {378},
  doi = {10.1038/s41597-023-02286-w}
}

@misc{lin2025somatomotor,
  author = {Lin, Fa-Hsuan and Foxe von Pechmann, Deirdre and Lankinen, Kaisu and Ahlfors, Seppo and Rosen, Bruce and Ahveninen, Jyrki and Hämäläinen, Matti and Raij, Tommi},
  title = {{somatomotor}},
  howpublished = {OpenNeuro [Data set]},
  year = {2025},
  url = {https://doi.org/10.18112/openneuro.ds006035.v1.0.0}
}

@article{nastase2021narratives,
  author = {Nastase, Samuel A. and Liu, Yun-Fei and Hillman, Hanna and Zadbood, Asieh and Hasenfratz, Liat and Keshavarzian, Neggin and Chen, Janice and Honey, Christopher J. and Yeshurun, Yaara and Regev, Mor and Nguyen, Mai and Chang, Claire H. C. and Baldassano, Christopher and Lositsky, Olga and Simony, Erez and Chow, Michael A. and Leong, Yuan Chang and Brooks, Paula P. and Micciche, Emily and Choe, Gina and Goldstein, Ariel and Vanderwal, Tamara and Halchenko, Yaroslav O. and Norman, Kenneth A. and Hasson, Uri},
  title = {The ``{Narratives}'' {fMRI} dataset for evaluating models of naturalistic language comprehension},
  year = {2021},
  journal = {Sci. Data},
  volume = {8},
  number = {1},
  pages = {250},
  doi = {10.1038/s41597-021-01033-3}
}

@article{bhattasali2020alice,
  author = {Bhattasali, Shohini and Brennan, Jonathan and Luh, Wen-Ming and Franzluebbers, Berta and Hale, John},
  title = {The {Alice Datasets}: {fMRI} \& {EEG} observations of natural language comprehension},
  journal = {Proceedings of the Twelfth Language Resources and Evaluation Conference},
  year = {2020},
  pages = {120--125},
  address = {Marseille, France},
  publisher = {European Language Resources Association},
  url = {https://aclanthology.org/2020.lrec-1.15/}
}

@article{xiao2025brainomni,
  author = {Xiao, Qinfan and Cui, Ziyun and Zhang, Chi and Chen, Siqi and Wu, Wen and Thwaites, Andrew and Woolgar, Alexandra and Zhou, Bowen and Zhang, Chao},
  title = {{BrainOmni}: A Brain Foundation Model for Unified {EEG} and {MEG} Signals},
  journal = {Advances in Neural Information Processing Systems},
  year = {2025},
  volume = {38},
  pages = {46081--46114},
  publisher = {Curran Associates, Inc.},
  doi = {10.52202/085713-1375}
}

@article{lu2025brain,
  author = {Lu, Wentao and Nie, Dong and Xue, Pengcheng and Cui, Zheng and Li, Piji and Zhang, Daoqiang and Wen, Xuyun},
  title = {Brain-Inspired {fMRI}-to-Text Decoding via Incremental and Wrap-Up Language Modeling},
  journal = {Advances in Neural Information Processing Systems},
  year = {2025},
  volume = {38},
  pages = {166540--166563},
  publisher = {Curran Associates, Inc.},
  doi = {10.52202/085713-5015}
}

@article{peterson2022ajile12,
  author = {Peterson, Steven M. and Singh, Satpreet H. and Dichter, Benjamin and Scheid, Michael and Rao, Rajesh P. N. and Brunton, Bingni W.},
  title = {{AJILE12}: Long-term naturalistic human intracranial neural recordings and pose},
  journal = {Sci. Data},
  year = {2022},
  volume = {9},
  number = {1},
  pages = {184},
  doi = {10.1038/s41597-022-01280-y}
}

@article{dubois2024reliability,
  author = {Dubois, Julien and Field, Ryan M. and Jawhar, Sami and Koch, Erin M. and Aghajan, Zahra M. and Miller, Naomi and Perdue, Katherine L. and Taylor, Moriah},
  title = {Reliability of brain metrics derived from a time-domain functional near-infrared spectroscopy system},
  journal = {Sci. Rep.},
  year = {2024},
  volume = {14},
  number = {1},
  pages = {17500},
  doi = {10.1038/s41598-024-68555-9}
}

@misc{duwadi2026cocktail,
  author = {Duwadi, Sudan and Rogers, De'Ja and Boyd, Alex D. and Carlton, Laura B. and Zhang, Yiwen and Kawai Gaona, Anna and Pathiyaparambath, Aneesa Diya and Chaudhury, Ravin and Magee, Grace and Zimmermann, Bernhard B. and O'Brien, Walker J. and von Lühmann, Alexander and Boas, David A. and Yücel, Meryem A. and Sen, Kamal},
  title = {Decoding spatial attention in the cocktail party problem using wearable whole-head high-density {fNIRS}},
  howpublished = {bioRxiv 2026.05.06.722322 [Preprint] (2026). \url{https://doi.org/10.64898/2026.05.06.722322}}
}

@article{wang2022fnirst,
  author = {Wang, Zenghui and Zhang, Jun and Zhang, Xiaochu and Chen, Peng and Wang, Bing},
  title = {Transformer model for functional near-infrared spectroscopy classification},
  journal = {IEEE J. Biomed. Health Inform.},
  year = {2022},
  volume = {26},
  number = {6},
  pages = {2559--2569},
  doi = {10.1109/JBHI.2022.3140531}
}

@misc{feczko2021abcd,
  author = {Feczko, Eric and Conan, Greg and Marek, Scott and Tervo-Clemmens, Brenden and Cordova, Michaela and Doyle, Olivia and Earl, Eric and Perrone, Anders and Sturgeon, Darrick and Klein, Rachel and Harman, Gareth and Kilamovich, Dakota and Hermosillo, Robert and Miranda-Dominguez, Oscar and Adebimpe, Azeez and Bertolero, Maxwell and Cieslak, Matthew and Covitz, Sydney and Hendrickson, Timothy and Juliano, Anthony C. and Snider, Kathy and Moore, Lucille A. and Uriartel, Johnny and Graham, Alice M. and Calabro, Finn and Rosenberg, Monica D. and Rapuano, Kristina M. and Casey, BJ and Watts, Richard and Hagler, Donald and Thompson, Wesley K. and Nichols, Thomas E. and Hoffman, Elizabeth and Luna, Beatriz and Garavan, Hugh and Satterthwaite, Theodore D. and Ewing, Sarah Feldstein and Nagel, Bonnie and Dosenbach, Nico U.F. and Fair, Damien A.},
  title = {Adolescent Brain Cognitive Development ({ABCD}) Community {MRI} Collection and Utilities},
  howpublished = {bioRxiv 2021.07.09.451638 [Preprint] (2021). \url{https://doi.org/10.1101/2021.07.09.451638}}
}

@article{glasser2013minimal,
  author = {Glasser, Matthew F. and Sotiropoulos, Stamatios N. and Wilson, J. Anthony and Coalson, Timothy S. and Fischl, Bruce and Andersson, Jesper L. and Xu, Junqian and Jbabdi, Saad and Webster, Matthew and Polimeni, Jonathan R. and Van Essen, David C. and Jenkinson, Mark},
  title = {The minimal preprocessing pipelines for the Human Connectome Project},
  journal = {NeuroImage},
  year = {2013},
  volume = {80},
  pages = {105--124},
  doi = {10.1016/j.neuroimage.2013.04.127}
}

@article{schaefer2018local,
  author = {Schaefer, Alexander and Kong, Ru and Gordon, Evan M. and Laumann, Timothy O. and Zuo, Xi-Nian and Holmes, Avram J. and Eickhoff, Simon B. and Yeo, B. T. Thomas},
  title = {Local--Global Parcellation of the Human Cerebral Cortex from Intrinsic Functional Connectivity {MRI}},
  journal = {Cereb. Cortex},
  year = {2018},
  volume = {28},
  number = {9},
  pages = {3095--3114},
  doi = {10.1093/cercor/bhx179}
}

@article{yeo2011organization,
  author = {Yeo, B. T. Thomas and Krienen, Fenna M. and Sepulcre, Jorge and Sabuncu, Mert R. and Lashkari, Danial and Hollinshead, Marisa and Roffman, Joshua L. and Smoller, Jordan W. and Zollei, Lilla and Polimeni, Jonathan R. and Fischl, Bruce and Liu, Hesheng and Buckner, Randy L.},
  title = {The Organization of the Human Cerebral Cortex Estimated by Intrinsic Functional Connectivity},
  journal = {J. Neurophysiol.},
  year = {2011},
  volume = {106},
  number = {3},
  pages = {1125--1165},
  doi = {10.1152/jn.00338.2011}
}

@article{chen2023faced,
  author  = {Chen, Jingjing and Wang, Xiaobin and Huang, Chen and Hu, Xin and Shen, Xinke and Zhang, Dan},
  title   = {A Large Finer-grained Affective Computing {EEG} Dataset},
  journal = {Sci. Data},
  year    = {2023},
  volume  = {10},
  number  = {1},
  pages   = {740},
  doi     = {10.1038/s41597-023-02650-w}
}

@article{liu2022seedv,
  author  = {Liu, Wei and Qiu, Jie-Lin and Zheng, Wei-Long and Lu, Bao-Liang},
  title   = {Comparing Recognition Performance and Robustness of Multimodal Deep Learning Models for Multimodal Emotion Recognition},
  journal = {IEEE Trans. Cogn. Dev. Syst.},
  year    = {2022},
  volume  = {14},
  number  = {2},
  pages   = {715--729},
  doi     = {10.1109/TCDS.2021.3071170}
}

@misc{brunner2008bcic2a,
  author      = {Brunner, Clemens and Leeb, Robert and M{\"u}ller-Putz, Gernot R. and Schl{\"o}gl, Alois and Pfurtscheller, Gert},
  title       = {{BCI} Competition 2008---{Graz} data set {A}},
  year        = {2008},
  note        = {\url{https://www.bbci.de/competition/iv/desc_2a.pdf}}
}

@article{lioi2020simultaneous,
  author  = {Lioi, Giulia and Cury, Claire and Perronnet, Lorraine and Mano, Marsel and Bannier, Elise and L{\'e}cuyer, Anatole and Barillot, Christian},
  title   = {Simultaneous {EEG}-{fMRI} during a Neurofeedback Task, a Brain Imaging Dataset for Multimodal Data Integration},
  journal = {Sci. Data},
  year    = {2020},
  volume  = {7},
  number  = {1},
  pages   = {173},
  doi     = {10.1038/s41597-020-0498-3}
}

@article{ma2022shumi,
  author  = {Ma, Jun and Yang, Banghua and Qiu, Wenzheng and Li, Yunzhe and Gao, Shouwei and Xia, Xinxing},
  title   = {A Large {EEG} Dataset for Studying Cross-session Variability in Motor Imagery Brain--Computer Interface},
  journal = {Sci. Data},
  year    = {2022},
  volume  = {9},
  number  = {1},
  pages   = {531},
  doi     = {10.1038/s41597-022-01647-1}
}

@article{golmohammadi2019automatic,
  author  = {Golmohammadi, Meysam and Harati Nejad Torbati, Amir Hossein and Lopez de Diego, Silvia and Obeid, Iyad and Picone, Joseph},
  title   = {Automatic Analysis of {EEGs} Using Big Data and Hybrid Deep Learning Architectures},
  journal = {Front. Hum. Neurosci.},
  year    = {2019},
  volume  = {13},
  pages   = {76},
  doi     = {10.3389/fnhum.2019.00076}
}

@article{angrick2024speech,
  author  = {Angrick, Miguel and Luo, Shiyu and Rabbani, Qinwan and Candrea, Daniel N. and Shah, Samyak and Milsap, Griffin W. and Anderson, William S. and Gordon, Chad R. and Rosenblatt, Kathryn R. and Clawson, Lora and Tippett, Donna C. and Maragakis, Nicholas and Tenore, Francesco V. and Fifer, Matthew S. and Hermansky, Hynek and Ramsey, Nick F. and Crone, Nathan E.},
  title   = {Online Speech Synthesis Using a Chronically Implanted Brain--Computer Interface in an Individual with {ALS}},
  journal = {Sci. Rep.},
  year    = {2024},
  volume  = {14},
  number  = {1},
  pages   = {9617},
  doi     = {10.1038/s41598-024-60277-2}
}

@article{wang2022smn4lang,
  author  = {Wang, Shaonan and Zhang, Xiaohan and Zhang, Jiajun and Zong, Chengqing},
  title   = {A Synchronized Multimodal Neuroimaging Dataset for Studying Brain Language Processing},
  journal = {Sci. Data},
  year    = {2022},
  volume  = {9},
  number  = {1},
  pages   = {590},
  doi     = {10.1038/s41597-022-01708-5}
}

@article{li2022lppc,
  author  = {Li, Jixing and Bhattasali, Shohini and Zhang, Shulin and Franzluebbers, Berta and Luh, Wen-Ming and Spreng, R. Nathan and Brennan, Jonathan R. and Yang, Yiming and Pallier, Christophe and Hale, John},
  title   = {{Le Petit Prince} Multilingual Naturalistic {fMRI} Corpus},
  journal = {Sci. Data},
  year    = {2022},
  volume  = {9},
  number  = {1},
  pages   = {530},
  doi     = {10.1038/s41597-022-01625-7}
}

@article{momenian2024lpphk,
  author  = {Momenian, Mohammad and Ma, Zhengwu and Wu, Shuyi and Wang, Chengcheng and Brennan, Jonathan and Hale, John and Meyer, Lars and Li, Jixing},
  title   = {{Le Petit Prince Hong Kong} ({LPPHK}): Naturalistic {fMRI} and {EEG} Data from Older {Cantonese} Speakers},
  journal = {Sci. Data},
  year    = {2024},
  volume  = {11},
  number  = {1},
  pages   = {992},
  doi     = {10.1038/s41597-024-03745-8}
}

@article{hanke2014studyforrest,
  author  = {Hanke, Michael and Baumgartner, Florian J. and Ibe, Pierre and Kaule, Falko R. and Pollmann, Stefan and Speck, Oliver and Zinke, Wolf and Stadler, J{\"o}rg},
  title   = {A High-resolution {7-Tesla} {fMRI} Dataset from Complex Natural Stimulation with an Audio Movie},
  journal = {Sci. Data},
  year    = {2014},
  volume  = {1},
  number  = {1},
  pages   = {140003},
  doi     = {10.1038/sdata.2014.3}
}

@article{gong2023nod,
  author  = {Gong, Zhengxin and Zhou, Ming and Dai, Yuxuan and Wen, Yushan and Liu, Youyi and Zhen, Zonglei},
  title   = {A Large-scale {fMRI} Dataset for the Visual Processing of Naturalistic Scenes},
  journal = {Sci. Data},
  year    = {2023},
  volume  = {10},
  number  = {1},
  pages   = {559},
  doi     = {10.1038/s41597-023-02471-x}
}

@article{allen2022nsd,
  author  = {Allen, Emily J. and St-Yves, Ghislain and Wu, Yihan and Breedlove, Jesse L. and Prince, Jacob S. and Dowdle, Logan T. and Nau, Matthias and Caron, Brad and Pestilli, Franco and Charest, Ian and Hutchinson, J. Benjamin and Naselaris, Thomas and Kay, Kendrick},
  title   = {A Massive {7T} {fMRI} Dataset to Bridge Cognitive Neuroscience and Artificial Intelligence},
  journal = {Nat. Neurosci.},
  year    = {2022},
  volume  = {25},
  number  = {1},
  pages   = {116--126},
  doi     = {10.1038/s41593-021-00962-x}
}

@article{jung2025spacetop,
  author  = {Jung, Heejung and Amini, Maryam and Hunt, Bethany J. and Murphy, Eilis I. and Sadil, Patrick and Halchenko, Yaroslav O. and Petre, Bogdan and Miao, Zizhuang and Kragel, Philip A. and Han, Xiaochun and Heilicher, Mickela O. and Sun, Michael and Collins, Owen G. and Lindquist, Martin A. and Wager, Tor D.},
  title   = {Spacetop: A Multimodal {fMRI} Dataset Unifying Naturalistic Processes with a Rich Array of Experimental Tasks},
  journal = {Sci. Data},
  year    = {2025},
  volume  = {12},
  number  = {1},
  pages   = {1465},
  doi     = {10.1038/s41597-025-05154-x}
}

@article{dupre2016autobiographical,
  author  = {DuPre, Elizabeth and Luh, Wen-Ming and Spreng, R. Nathan},
  title   = {Multi-echo {fMRI} replication sample of autobiographical memory, prospection and theory of mind reasoning tasks},
  journal = {Sci. Data},
  year    = {2016},
  volume  = {3},
  number  = {1},
  pages   = {160116},
  doi     = {10.1038/sdata.2016.116}
}

@article{lee2021imagined,
  author  = {Lee, Sangil and Parthasarathi, Trishala and Kable, Joseph W.},
  title   = {The Ventral and Dorsal Default Mode Networks Are Dissociably Modulated by the Vividness and Valence of Imagined Events},
  journal = {J. Neurosci.},
  year    = {2021},
  volume  = {41},
  number  = {24},
  pages   = {5243--5250},
  doi     = {10.1523/JNEUROSCI.1273-20.2021}
}

@article{wang2025neuralcode,
  author  = {Wang, Haibao and Ho, Jun Kai and Cheng, Fan L. and Aoki, Shuntaro C. and Muraki, Yusuke and Tanaka, Misato and Park, Jong-Yun and Kamitani, Yukiyasu},
  title   = {Inter-individual and inter-site neural code conversion without shared stimuli},
  journal = {Nat. Comput. Sci.},
  year    = {2025},
  volume  = {5},
  number  = {7},
  pages   = {534--546},
  doi     = {10.1038/s43588-025-00826-5}
}

@article{jiang2024labram,
  author  = {Jiang, Wei-Bang and Zhao, Li-Ming and Lu, Bao-Liang},
  title   = {Large Brain Model for Learning Generic Representations with Tremendous {EEG} Data in {BCI}},
  journal = {The Twelfth International Conference on Learning Representations},
  year    = {2024},
  pages   = {35429--35450},
  url     = {https://openreview.net/forum?id=QzTpTRVtrP}
}

@article{wang2025cbramod,
  author  = {Wang, Jiquan and Zhao, Sha and Luo, Zhiling and Zhou, Yangxuan and Jiang, Haiteng and Li, Shijian and Li, Tao and Pan, Gang},
  title   = {{CBraMod}: A Criss-Cross Brain Foundation Model for {EEG} Decoding},
  journal = {The Thirteenth International Conference on Learning Representations},
  year    = {2025},
  pages   = {62056--62092},
  url     = {https://openreview.net/forum?id=NPNUHgHF2w}
}

@article{jing2023sparcnet,
  author  = {Jing, Jin and Ge, Wendong and Hong, Shenda and Fernandes, Marta Bento and Lin, Zhen and Yang, Chaoqi and An, Sungtae and Struck, Aaron F and Herlopian, Aline and Karakis, Ioannis and others},
  title   = {Development of Expert-Level Classification of Seizures and Rhythmic and Periodic Patterns During {EEG} Interpretation},
  journal = {Neurology},
  year    = {2023},
  volume  = {100},
  number  = {17},
  pages   = {e1750--e1762},
  doi     = {10.1212/WNL.0000000000207127}
}

@article{wang2026neurostorm,
  author  = {Wang, Cheng and Jiang, Yu and Peng, Zhihao and Li, Chenxin and Bang, Chang-Bae and Zhao, Lin and Fu, Wanyi and Lv, Jinglei and Sepulcre, Jorge and Yang, Carl and others},
  title   = {Towards a General-Purpose Foundation Model for Functional {MRI} Analysis},
  journal = {Nat. Biomed. Eng.},
  year    = {2026},
  note     = {\url{https://doi.org/10.1038/s41551-026-01666-y}}
}

@article{chen2024bgem3,
  author       = {Chen, Jianlyu and Xiao, Shitao and Zhang, Peitian and Luo, Kun and Lian, Defu and Liu, Zheng},
  title        = {{M3-Embedding}: Multi-Linguality, Multi-Functionality, Multi-Granularity Text Embeddings Through Self-Knowledge Distillation},
  journal    = {Findings of the Association for Computational Linguistics: ACL 2024},
  year         = {2024},
  pages        = {2318--2335},
  address      = {Bangkok, Thailand},
  publisher    = {Association for Computational Linguistics},
  doi          = {10.18653/v1/2024.findings-acl.137},
  url          = {https://aclanthology.org/2024.findings-acl.137/}
}

@misc{abouelenin2025phi4mini,
  author       = {Abouelenin, Abdelrahman and Ashfaq, Atabak and Atkinson, Adam and Awadalla, Hany and Bach, Nguyen and Bao, Jianmin and Benhaim, Alon and Cai, Martin and Chaudhary, Vishrav and Chen, Congcong and others},
  title        = {{Phi-4-Mini} Technical Report: Compact Yet Powerful Multimodal Language Models via Mixture-of-{LoRA}s},
  howpublished = {arXiv:2503.01743 [cs.CL]},
  year         = {2025}
}

@article{li2023blip2,
  author  = {Li, Junnan and Li, Dongxu and Savarese, Silvio and Hoi, Steven},
  title   = {{BLIP}-2: Bootstrapping Language-Image Pre-training with Frozen Image Encoders and Large Language Models},
  journal = {Proceedings of the 40th International Conference on Machine Learning},
  year    = {2023},
  volume  = {202},
  series  = {Proceedings of Machine Learning Research},
  pages   = {19730--19742},
  publisher = {PMLR},
  url     = {https://proceedings.mlr.press/v202/li23q.html}
}

@article{meng2024pissa,
  author  = {Meng, Fanxu and Wang, Zhaohui and Zhang, Muhan},
  title   = {{PiSSA}: Principal Singular Values and Singular Vectors Adaptation of Large Language Models},
  journal = {Advances in Neural Information Processing Systems},
  year    = {2024},
  volume  = {37},
  pages   = {121038--121072},
  publisher = {Curran Associates, Inc.},
  doi     = {10.52202/079017-3846}
}

@article{radford2021clip,
  author  = {Radford, Alec and Kim, Jong Wook and Hallacy, Chris and Ramesh, Aditya and Goh, Gabriel and Agarwal, Sandhini and Sastry, Girish and Askell, Amanda and Mishkin, Pamela and Clark, Jack and Krueger, Gretchen and Sutskever, Ilya},
  title   = {Learning Transferable Visual Models from Natural Language Supervision},
  journal = {Proceedings of the 38th International Conference on Machine Learning},
  year    = {2021},
  volume  = {139},
  series  = {Proceedings of Machine Learning Research},
  pages   = {8748--8763},
  publisher = {PMLR},
  url     = {https://proceedings.mlr.press/v139/radford21a.html}
}

@article{rombach2022ldm,
  author  = {Rombach, Robin and Blattmann, Andreas and Lorenz, Dominik and Esser, Patrick and Ommer, Bj{\"o}rn},
  title   = {High-Resolution Image Synthesis with Latent Diffusion Models},
  journal = {2022 IEEE/CVF Conference on Computer Vision and Pattern Recognition},
  year    = {2022},
  pages   = {10674--10685},
  publisher = {IEEE},
  doi     = {10.1109/CVPR52688.2022.01042}
}

@article{song2021ddim,
  author  = {Song, Jiaming and Meng, Chenlin and Ermon, Stefano},
  title   = {Denoising Diffusion Implicit Models},
  journal = {The Ninth International Conference on Learning Representations},
  year    = {2021},
  url     = {https://openreview.net/forum?id=St1giarCHLP}
}

@article{hinton2002poe,
  author  = {Hinton, Geoffrey E.},
  title   = {Training Products of Experts by Minimizing Contrastive Divergence},
  journal = {Neural Computation},
  year    = {2002},
  volume  = {14},
  number  = {8},
  pages   = {1771--1800},
  doi     = {10.1162/089976602760128018}
}

@article{bethlehem2022braincharts,
  author  = {Bethlehem, Richard AI and Seidlitz, Jakob and White, Simon R and Vogel, Jacob W and Anderson, Kevin M and Adamson, Chris and Adler, Sophie and Alexopoulos, George S and Anagnostou, Evdokia and Areces-Gonzalez, Ariosky and others},
  title   = {Brain charts for the human lifespan},
  journal = {Nature},
  year    = {2022},
  volume  = {604},
  number  = {7906},
  pages   = {525--533},
  doi     = {10.1038/s41586-022-04554-y}
}

@article{Tyree2023science,
author = {Timothy J. Tyree  and Michael Metke  and Cory T. Miller },
title = {Cross-modal representation of identity in the primate hippocampus},
journal = {Science},
volume = {382},
number = {6669},
pages = {417-423},
year = {2023},
doi = {10.1126/science.adf0460},

}

@article{Dijksterhuis2024science,
author = {D. E. Dijksterhuis  and M. W. Self  and J. K. Possel  and J. C. Peters  and E. C. W. van Straaten  and S. Idema  and J. C. Baaijen  and S. M. A. van der Salm  and E. J. Aarnoutse  and N. C. E. van Klink  and P. van Eijsden  and S. Hanslmayr  and R. Chelvarajah  and F. Roux  and L. D. Kolibius  and V. Sawlani  and D. T. Rollings  and S. Dehaene  and P. R. Roelfsema },
title = {Pronouns reactivate conceptual representations in human hippocampal neurons},
journal = {Science},
volume = {385},
number = {6716},
pages = {1478-1484},
year = {2024},
doi = {10.1126/science.adr2813},

}

@article{Can2023science,
author = {Can Liu  and Ralitsa Todorova  and Wenbo Tang  and Azahara Oliva  and Antonio Fernandez-Ruiz },
title = {Associative and predictive hippocampal codes support memory-guided behaviors},
journal = {Science},
volume = {382},
number = {6668},
pages = {eadi8237},
year = {2023},
doi = {10.1126/science.adi8237},

}

@article{Walter2019science,
author = {Walter G. Gonzalez  and Hanwen Zhang  and Anna Harutyunyan  and Carlos Lois },
title = {Persistence of neuronal representations through time and damage in the hippocampus},
journal = {Science},
volume = {365},
number = {6455},
pages = {821-825},
year = {2019},
doi = {10.1126/science.aav9199},

}

\clearpage
\renewcommand{\thefigure}{\arabic{figure}}
\renewcommand{\theHfigure}{ED.\arabic{figure}}
\renewcommand{\thetable}{\arabic{table}}
\renewcommand{\theHtable}{ED.\arabic{table}}
\renewcommand{\tablename}{Extended Data Table}
\setcounter{figure}{0}
\setcounter{table}{0}

\begin{figure*}[p]
\centering
\small\linespread{1.0}\selectfont
\refstepcounter{figure}\label{fig:ed-smn4lang-cortical}
\sbox{\mainfigurelegend}{\parbox{\textwidth}{\noindent\textbf{Extended Data Fig. \thefigure\quad Semantic-attribution maps across participants.} Maps of the low-level neural observation representation and high-level meta-neural semantic representation are shown for 12 SMN4Lang participants. Percentages indicate cortical coverage.\par}}
\includegraphics[width=\textwidth,height=\dimexpr0.96\textheight-\ht\mainfigurelegend-\dp\mainfigurelegend-10pt\relax,keepaspectratio]{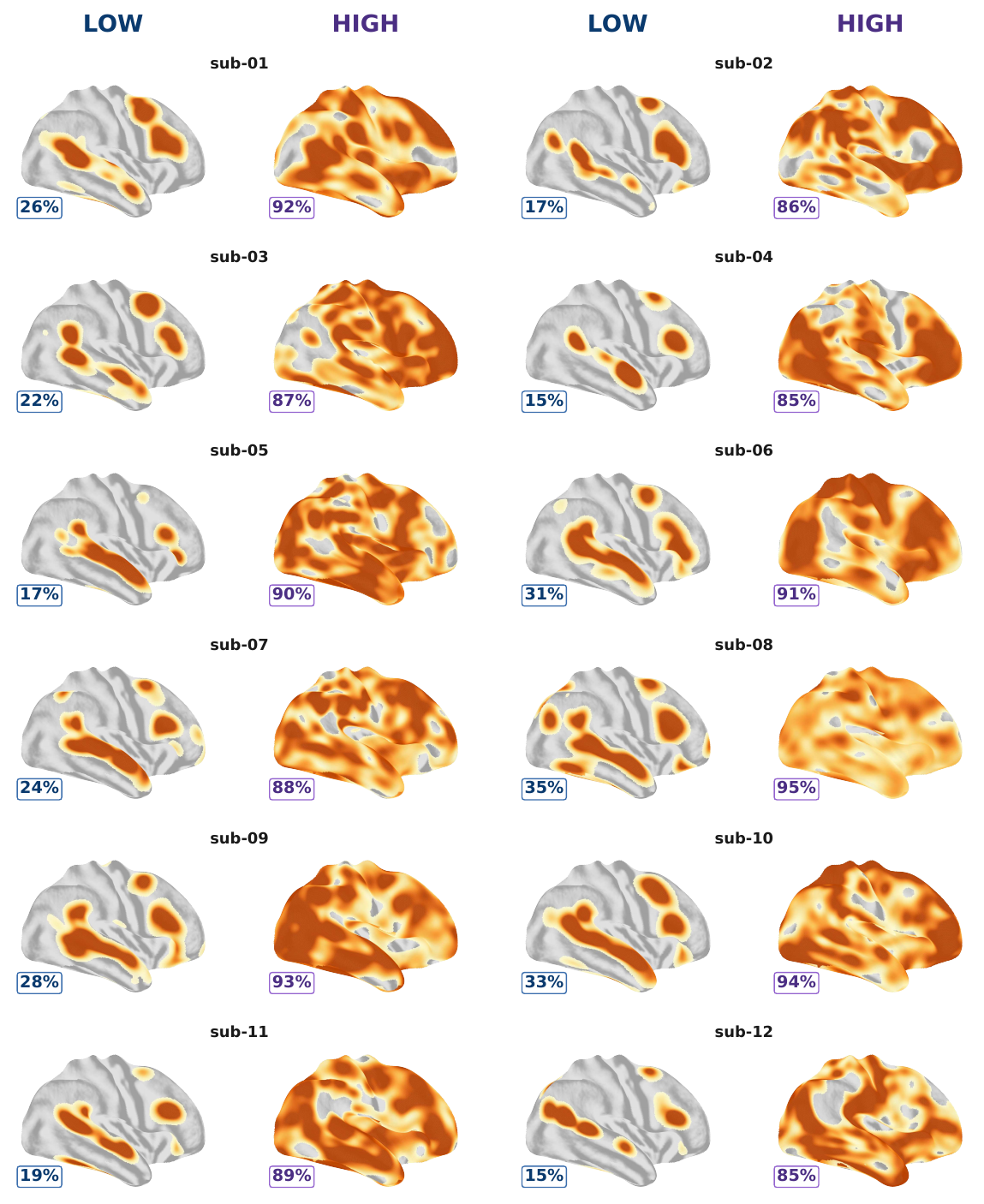}\par
\vspace{6pt}
\usebox{\mainfigurelegend}
\end{figure*}
\clearpage

\begin{figure*}[p]
\centering
\small\linespread{1.0}\selectfont
\refstepcounter{figure}\label{fig:ed-rating}
\sbox{\mainfigurelegend}{\parbox{\textwidth}{\noindent\textbf{Extended Data Fig. \thefigure\quad Subjective-state prediction across Spacetop tasks.} \textbf{a},~ Prediction--rating correlations for expectation intensity, expectation valence, feeling intensity and feeling valence in Spacetop Narratives, and for likeability, mentalizing and similarity in Spacetop Shortvideo. Each target included $n=30$ participants. \textbf{b},~ Prediction of participant-specific rating residuals after removal of the shared stimulus-locked component. Green denotes NeuroSTORM and red denotes the proposed method. Box plots show the median (center line), 25th and 75th percentiles (box limits), and 1.5 $\times$ IQR whiskers; values outside the whiskers are shown individually. Violin plots show participant-level distributions. Correlation coefficients were Fisher $z$-transformed before statistical testing; raw Pearson $r$ values are shown. Two-way repeated-measures ANOVA followed by \v{S}id\'{a}k-adjusted multiple comparisons was used in each task, with comparisons between methods at each rating dimension. Asterisks denote statistical significance (*$P < 0.05$, **$P < 0.01$, ***$P < 0.001$ and ****$P < 0.0001$).\par}}
\includegraphics[width=\textwidth,height=\dimexpr0.96\textheight-\ht\mainfigurelegend-\dp\mainfigurelegend-10pt\relax,keepaspectratio]{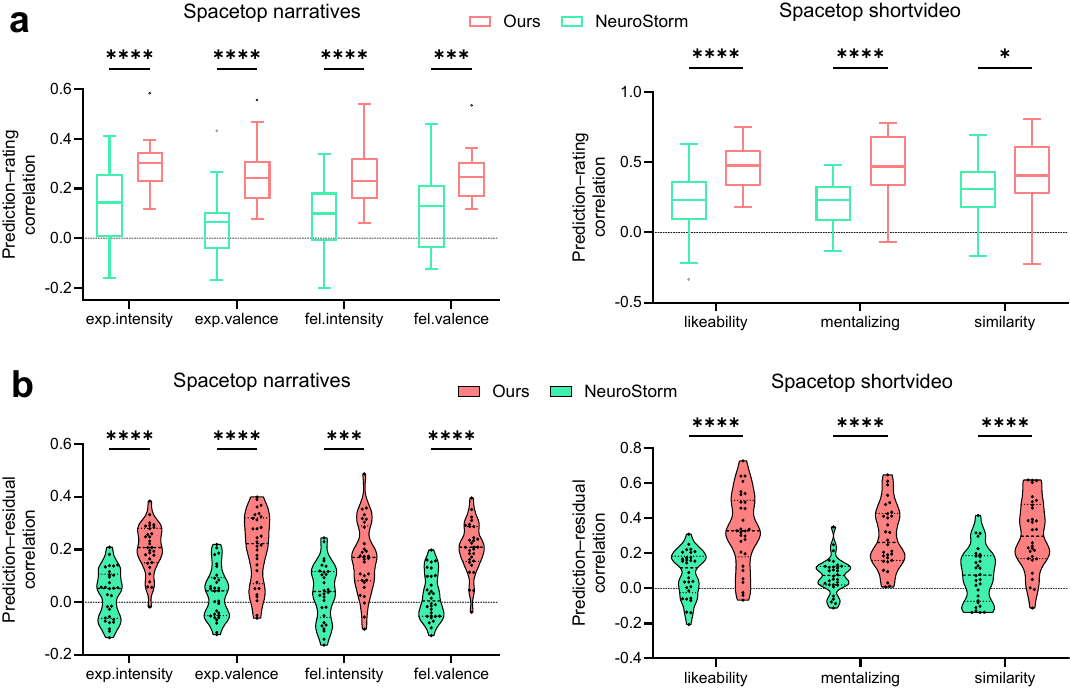}\par
\vspace{6pt}
\usebox{\mainfigurelegend}
\end{figure*}
\clearpage

\begin{figure*}[p]
\centering
\small\linespread{1.0}\selectfont
\refstepcounter{figure}\label{fig:ed-clustering}
\sbox{\mainfigurelegend}{\parbox{\textwidth}{\noindent\textbf{Extended Data Fig. \thefigure\quad Representational geometry of low-level and high-level representations across cognitive categories.} \textbf{a},~ Multidimensional-scaling visualizations for motor, perception and internal-mentation categories in the low-level neural observation and high-level meta-neural semantic representations. \textbf{b},~ Within-class dispersion, between-class separation and category-structure representational similarity analysis (RSA) for the two representations. Violin plots show distributions, with internal lines indicating medians and quartiles. Asterisks denote statistical significance (*$P < 0.05$, **$P < 0.01$, ***$P < 0.001$ and ****$P < 0.0001$).\par}}
\includegraphics[width=\textwidth,height=\dimexpr0.96\textheight-\ht\mainfigurelegend-\dp\mainfigurelegend-10pt\relax,keepaspectratio]{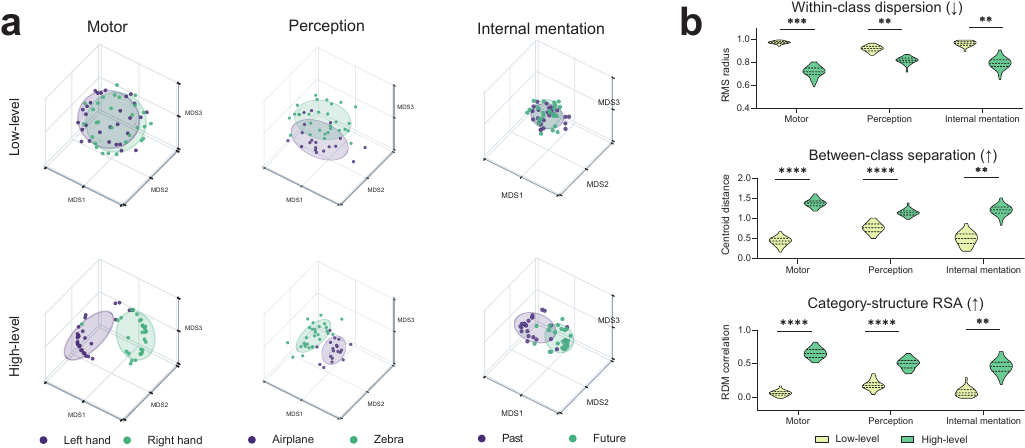}\par
\vspace{6pt}
\usebox{\mainfigurelegend}
\end{figure*}
\clearpage

\begin{figure*}[p]
\centering
\small\linespread{1.0}\selectfont
\refstepcounter{figure}\label{fig:ed-spacetop-rsa}
\sbox{\mainfigurelegend}{\parbox{\textwidth}{\noindent\textbf{Extended Data Fig. \thefigure\quad Neural representations relate to subjective ratings in Spacetop.} \textbf{a},~ Association between neural representational distance and rating distance across 136 participant pairs formed from $n=17$ participants in Narratives. \textbf{b},~ Participant-level RSA correlations between neural representations and Narratives ratings. \textbf{c},~ Corresponding participant-pair analysis for Shortvideo ratings. \textbf{d},~ Participant-level RSA correlations for Shortvideo ratings. Green denotes NeuroSTORM and red denotes the proposed framework. Points in \textbf{a} and \textbf{c} denote participant pairs, and lines show linear fits with Pearson correlation coefficients. Violin plots in \textbf{b} and \textbf{d} show participant-level distributions; internal lines indicate the median and quartiles. Two-tailed paired $t$-tests on Fisher $z$-transformed participant-level RSA coefficients were used in \textbf{b} and \textbf{d}. Asterisks denote statistical significance (*$P < 0.05$, **$P < 0.01$, ***$P < 0.001$ and ****$P < 0.0001$).\par}}
\includegraphics[width=\textwidth,height=\dimexpr0.96\textheight-\ht\mainfigurelegend-\dp\mainfigurelegend-10pt\relax,keepaspectratio]{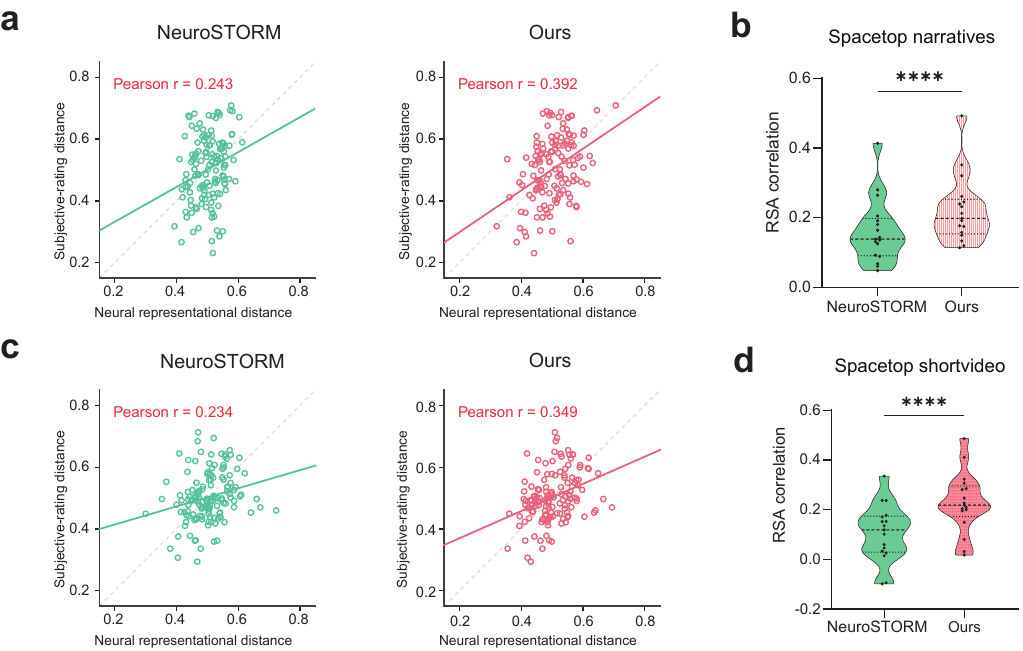}\par
\vspace{6pt}
\usebox{\mainfigurelegend}
\end{figure*}
\clearpage

\begin{figure*}[p]
\centering
\small\linespread{1.0}\selectfont
\refstepcounter{figure}\label{fig:ed-eigenmode-ablation}
\sbox{\mainfigurelegend}{\parbox{\textwidth}{\noindent\textbf{Extended Data Fig. \thefigure\quad Ablation analyses of cortical geometric eigenmodes.} \textbf{a},~ Within-participant decoding accuracy using Fourier, spherical-harmonic, graph-Laplacian and cortical geometric eigenmode bases. \textbf{b},~ Cross-participant decoding accuracy for the same bases. BCIC included $n=9$ participants, FACED $n=123$, SEED-V $n=16$ and Motor Imagery $n=10$. \textbf{c},~ Decoding accuracy as a function of the number of retained modes. Red boxes mark the dimensionalities used in the main analyses: 14, 20, 30 and 28 modes. Box plots show the median (center line), 25th and 75th percentiles (box limits), and 1.5 $\times$ IQR whiskers; values outside the whiskers are shown individually. One-way repeated-measures ANOVA with Greenhouse--Geisser correction followed by \v{S}id\'{a}k-adjusted multiple comparisons was used in \textbf{a} and \textbf{b}. Asterisks denote statistical significance (*$P < 0.05$, **$P < 0.01$, ***$P < 0.001$ and ****$P < 0.0001$).\par}}
\includegraphics[width=\textwidth,height=\dimexpr0.96\textheight-\ht\mainfigurelegend-\dp\mainfigurelegend-10pt\relax,keepaspectratio]{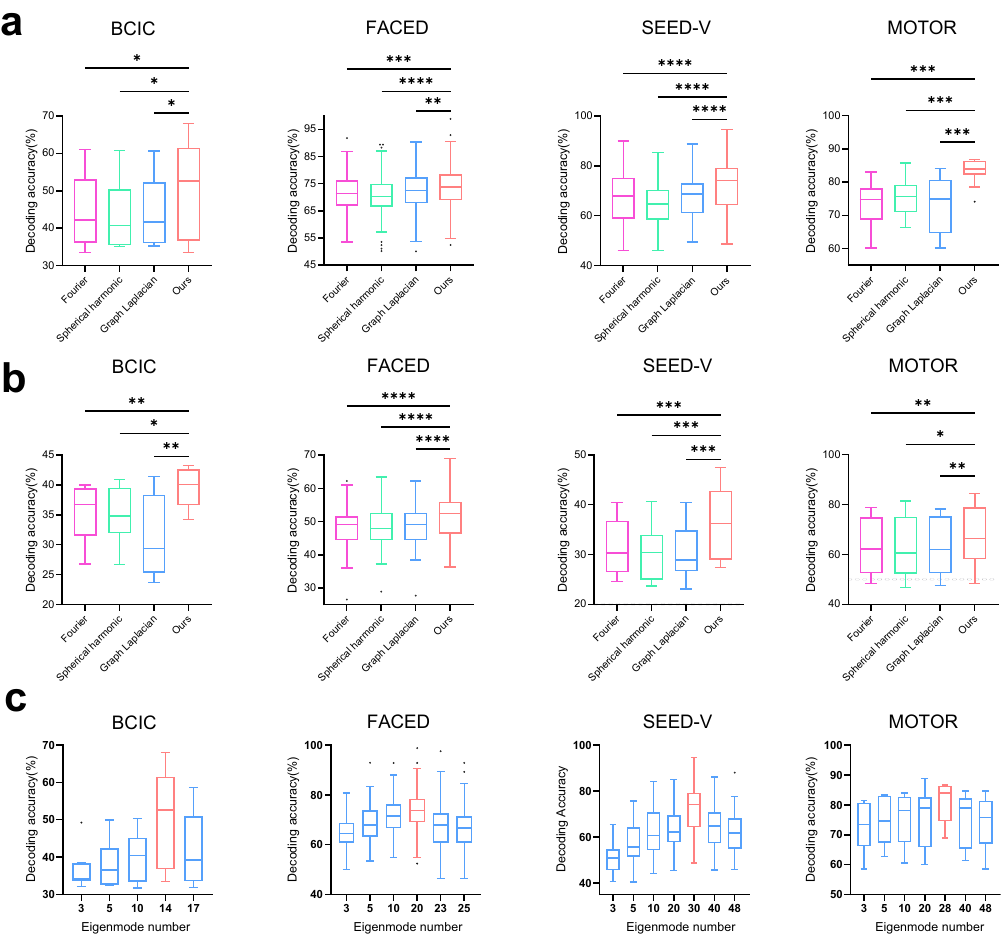}\par
\vspace{6pt}
\usebox{\mainfigurelegend}
\end{figure*}
\clearpage

\begin{figure*}[p]
\centering
\small\linespread{1.0}\selectfont
\refstepcounter{figure}\label{fig:ed-individual-geometry}
\sbox{\mainfigurelegend}{\parbox{\textwidth}{\noindent\textbf{Extended Data Fig. \thefigure\quad Individual cortical geometry and decoded features.} \textbf{a},~ Cortical surfaces and geometric eigenmodes 1, 5, 20, 40, 100, 200, 500 and 1,000 from the HCP S1200 group reference and three representative participants. \textbf{b--c},~ Group--individual correspondence across cortical eigenmode ranks. \textbf{b},~ Absolute same-rank spatial correlations $|r|$ between HCP S1200 group eigenmodes and individual eigenmodes for one participant from the Alice dataset. \textbf{c},~ Group--individual deviations in cortical-area-normalized eigenvalues across four eigenmode-rank bands in 26 Alice participants. For each participant, deviations were averaged within each band, and box plots summarize the participant distributions. Higher-rank eigenmodes showed weaker spatial correspondence but increasingly consistent eigenvalue scales. \textbf{d--f},~ Decoding accuracy using group-level and participant-specific cortical geometric eigenmodes in SomatoMotor (two-class; $n=5$), Motor Imagery (four-class; $n=10$) and the Imagined Speech OPM-MEG dataset collected for this study (three-class; $n=6$), respectively. \textbf{g--k},~ BGEScore using group-level and participant-specific cortical geometric eigenmodes in Narratives, Alice, LPPC-fMRI, StudyForrest and LPPC-HK, with $n=37$, 26, 112, 20 and 52 participants, respectively. Box plots in \textbf{c--k} show the median (center line), the 25th and 75th percentiles (box limits), and 1.5 $\times$ IQR whiskers; values outside the whiskers are shown individually. Two-tailed paired $t$-tests were used in \textbf{d--k}. Asterisks denote statistical significance (*$P < 0.05$, **$P < 0.01$, ***$P < 0.001$ and ****$P < 0.0001$).\par}}
\includegraphics[width=\textwidth,height=\dimexpr0.96\textheight-\ht\mainfigurelegend-\dp\mainfigurelegend-10pt\relax,keepaspectratio]{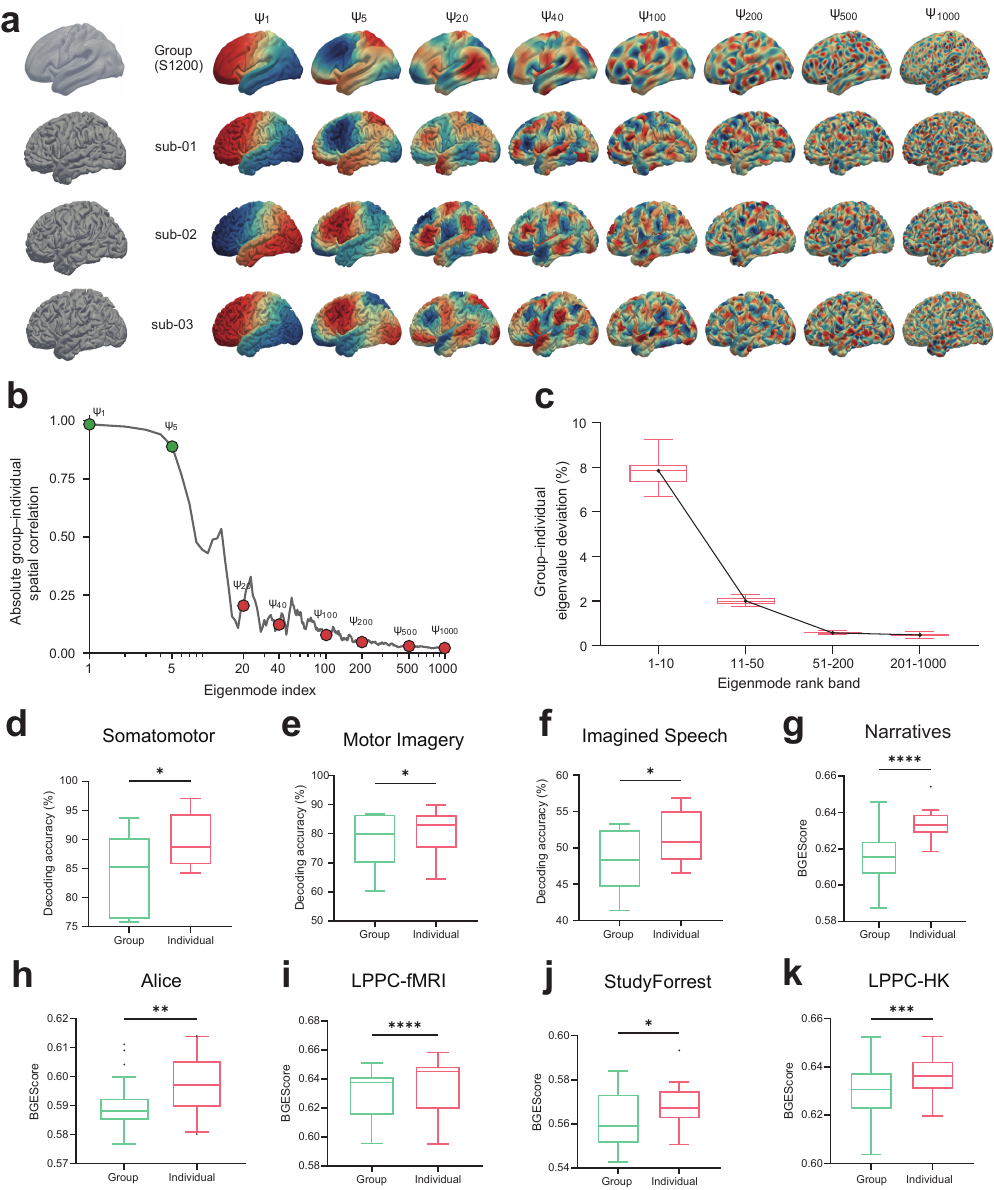}\par
\vspace{6pt}
\usebox{\mainfigurelegend}
\end{figure*}
\clearpage

\begin{figure*}[p]
\centering
\small\linespread{1.0}\selectfont
\refstepcounter{figure}\label{fig:ed-classification}
\sbox{\mainfigurelegend}{\parbox{\textwidth}{\noindent\textbf{Extended Data Fig. \thefigure\quad Additional discriminative decoding results.} \textbf{a},~ Inner-speech decoding accuracy across eight classes in Bimodal Inner Speech ($n=4$ participants) for LaBraM, BrainOmni, CBraMod and the proposed framework~\cite{simistiraliwicki2023bimodal}. \textbf{b},~ Somatomotor decoding accuracy across two classes in SomatoMotor ($n=5$) for the same methods. \textbf{c},~ Decoding accuracy across three classes in Dubois2024 ($n=49$)~\cite{dubois2024reliability} for the fNIRS Transformer~\cite{wang2022fnirst} and the proposed framework. \textbf{d},~ Two-class decoding accuracy for cocktail-party fNIRS; 30 of the 38 source participants had valid recordings for analysis~\cite{duwadi2026cocktail}. Box plots show the median (center line), 25th and 75th percentiles (box limits), and 1.5 $\times$ IQR whiskers; values outside the whiskers are shown individually. Brackets indicate pairwise comparisons. One-way repeated-measures ANOVA with Greenhouse--Geisser correction followed by \v{S}id\'{a}k-adjusted multiple comparisons was used in \textbf{a} and \textbf{b}. Two-sided paired $t$-tests were used in \textbf{c} and \textbf{d}. Asterisks denote statistical significance (*$P<0.05$, **$P<0.01$, ***$P<0.001$ and ****$P<0.0001$).\par}}
\includegraphics[width=\textwidth,height=\dimexpr0.96\textheight-\ht\mainfigurelegend-\dp\mainfigurelegend-10pt\relax,keepaspectratio]{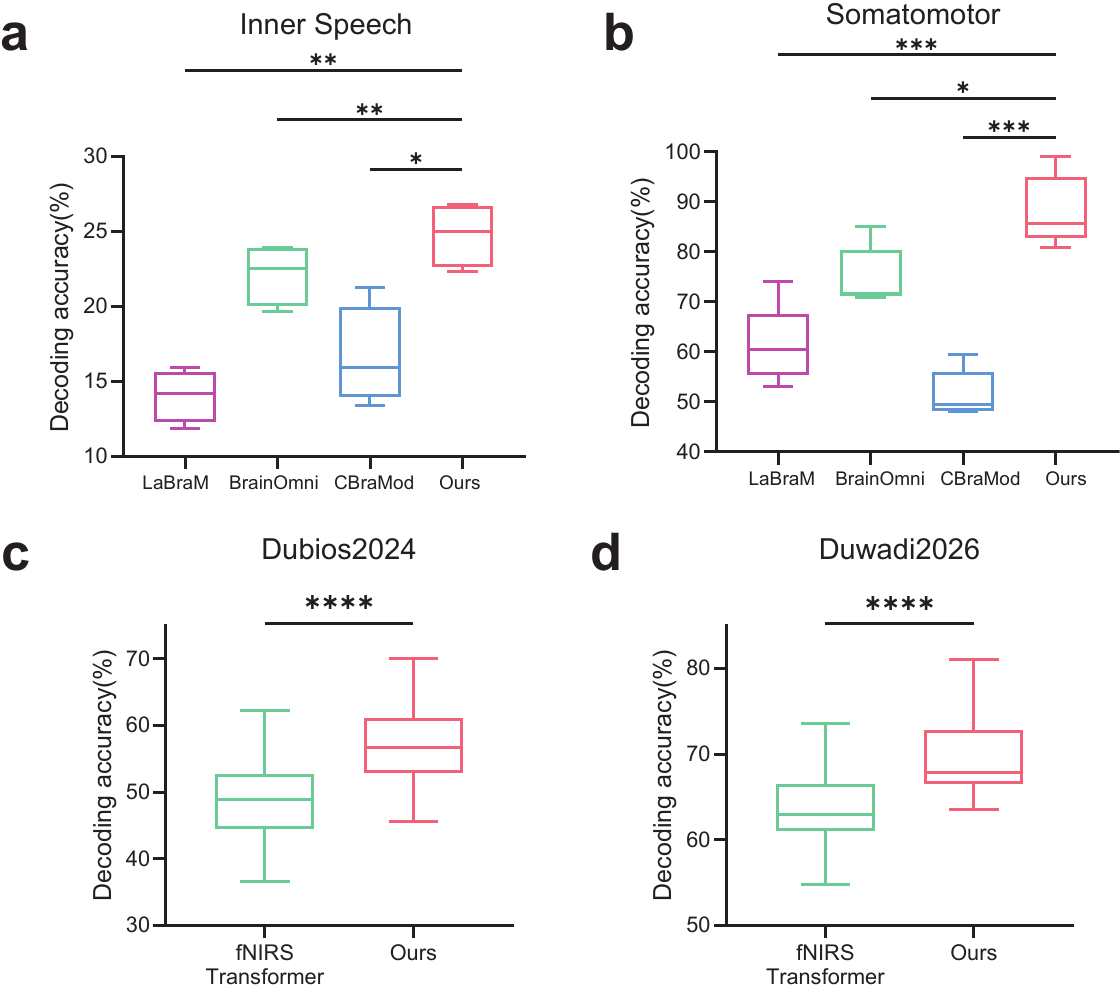}\par
\vspace{6pt}
\usebox{\mainfigurelegend}
\end{figure*}
\clearpage

\begin{figure*}[p]
\centering
\small\linespread{1.0}\selectfont
\refstepcounter{figure}\label{fig:ed-robustness}
\sbox{\mainfigurelegend}{\parbox{\textwidth}{\noindent\textbf{Extended Data Fig. \thefigure\quad Robustness of decoding performance.} \textbf{a},~ Decoding accuracy after adding increasing levels of noise relative to the training-fold channel standard deviation. \textbf{b},~ Accuracy after dropping increasing proportions of channels. \textbf{c},~ Accuracy as progressively smaller fractions of within-participant training data were used. \textbf{d},~ Accuracy in the cross-participant setting as increasing amounts of training data from the target participant were introduced. Each row compares BrainOmni with the proposed framework; columns show BCIC, FACED, SEED-V and Motor Imagery, with $n=9$, 123, 16 and 10 participants, respectively. Box plots show the median (center line), 25th and 75th percentiles (box limits), and 1.5 $\times$ IQR whiskers; values outside the whiskers are shown individually. Brackets indicate pairwise comparisons. Two-way repeated-measures ANOVA followed by \v{S}id\'{a}k-adjusted multiple comparisons was used for each dataset and analysis, with methods compared at each level. Asterisks denote statistical significance (*$P<0.05$, **$P<0.01$, ***$P<0.001$ and ****$P<0.0001$).\par}}
\includegraphics[width=\textwidth,height=\dimexpr0.96\textheight-\ht\mainfigurelegend-\dp\mainfigurelegend-10pt\relax,keepaspectratio]{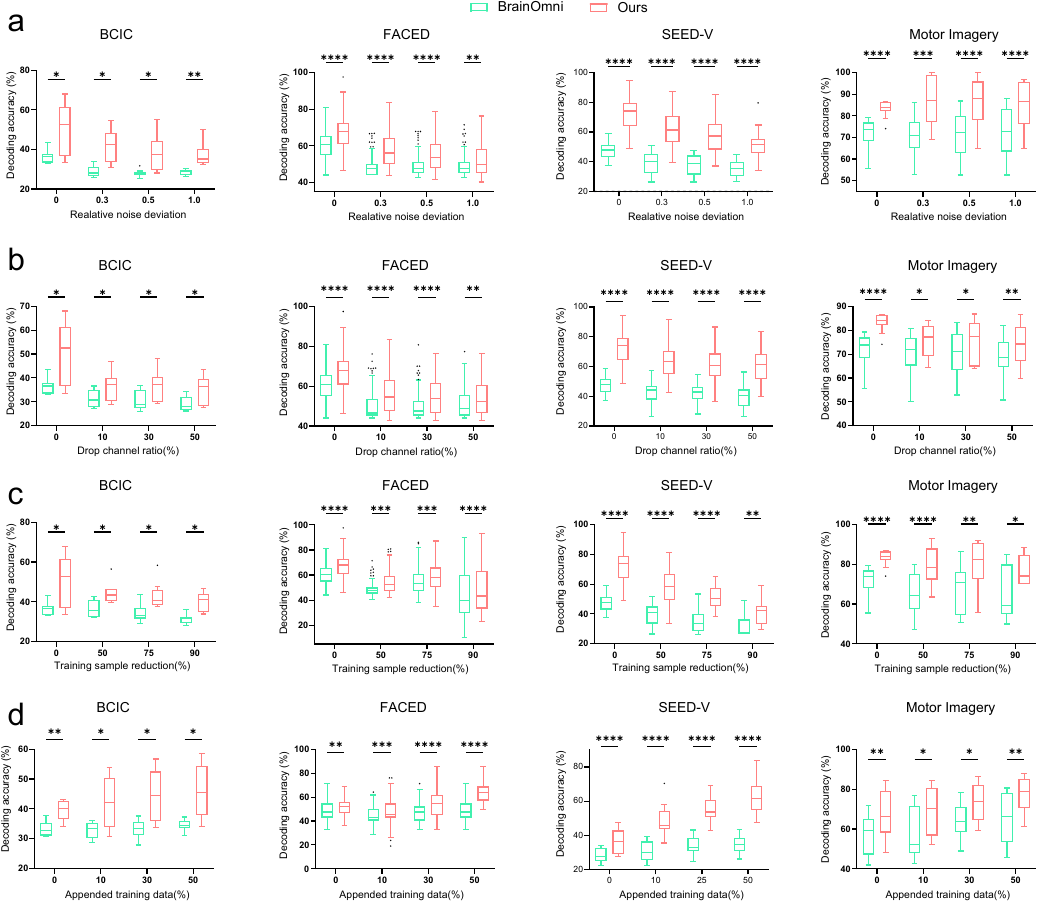}\par
\vspace{6pt}
\usebox{\mainfigurelegend}
\end{figure*}
\clearpage

\setcounter{table}{0}

\begin{table*}[!tbp]
\centering
\begin{minipage}{\textwidth}
\linespread{1.0}\selectfont
\caption{\textbf{Datasets used in this study.} $N$ denotes the reported cohort size (participants).}
\label{tab:datasets}
\small
\setlength{\tabcolsep}{3.0pt}
\renewcommand{\arraystretch}{1.10}
\begin{tabular*}{\textwidth}{@{\extracolsep{\fill}}cllclc@{}}
\hline
No. & Dataset & Modality & $N$ & Task & Ref. \\
\hline
1 & FACED & EEG & 123 & Affective video viewing & \cite{chen2023faced} \\
2 & SEED-V & EEG & 16 & Affective video viewing & \cite{liu2022seedv} \\
3 & BCIC-IV-2a & EEG & 9 & Motor imagery & \cite{brunner2008bcic2a} \\
4 & SHU-MI & EEG & 25 & Motor imagery & \cite{ma2022shumi} \\
5 & TUEV & EEG & 518 & Clinical EEG event classification & \cite{golmohammadi2019automatic} \\
6 & SomatoMotor & MEG + EEG & 5 & Somatosensory stimulation & \cite{lin2025somatomotor} \\
7 & Duwadi2026 & fNIRS & 38 & Auditory spatial attention & \cite{duwadi2026cocktail} \\
8 & Dubois2024 & fNIRS & 49 & Auditory and cognitive tasks & \cite{dubois2024reliability} \\
9 & Motor Imagery & fMRI + EEG & 10 & Motor imagery with neurofeedback & \cite{lioi2020simultaneous} \\
10 & Bimodal Inner Speech & fMRI + EEG & 4 & Inner speech production & \cite{simistiraliwicki2023bimodal} \\
11 & SMN4Lang & fMRI + MEG & 12 & Narrative listening & \cite{wang2022smn4lang} \\
12 & Narratives & fMRI & 345 & Narrative listening & \cite{nastase2021narratives} \\
13 & Alice & fMRI & 26 & Narrative listening & \cite{bhattasali2020alice} \\
14 & LPPC-fMRI & fMRI & 112 & Narrative listening & \cite{li2022lppc} \\
15 & LPPC-HK & fMRI + EEG & 52 & Narrative listening & \cite{momenian2024lpphk} \\
16 & StudyForrest & fMRI & 20 & Naturalistic movie viewing & \cite{hanke2014studyforrest} \\
17 & NOD & fMRI & 30 & Natural image viewing & \cite{gong2023nod} \\
18 & NSD & fMRI & 8 & Natural image viewing & \cite{allen2022nsd} \\
19 & Spacetop & fMRI & 101 & Affective and cognitive tasks & \cite{jung2025spacetop} \\
20 & DuPre2016 & fMRI & 31 & Autobiographical cognition & \cite{dupre2016autobiographical} \\
21 & Lee2021 & fMRI & 26 & Future event imagination & \cite{lee2021imagined} \\
22 & AJILE12 & ECoG & 12 & Naturalistic upper-limb movement & \cite{peterson2022ajile12} \\
23 & ECoG Speech & ECoG & 1 & Speech production & \cite{angrick2024speech} \\
\hline
\end{tabular*}
\end{minipage}
\end{table*}

\end{document}